\documentclass[review]{elsarticle}

\usepackage{hyperref}
\hypersetup{colorlinks,allcolors=black}

\journal{Computational Statistics \& Data Analysis}

\usepackage{amsmath,amsfonts,graphicx,natbib,psfrag,color}
\usepackage{algorithm}
\usepackage{booktabs}
\usepackage{natbib}
\usepackage{url}

\usepackage{multirow}
\usepackage{arydshln}

\newcommand{\indep}{\overset{indep}{\sim }}

\usepackage{changes}
\definechangesauthor[name=Umberto, color=blue]{up}

\begin{document}

\begin{frontmatter}

\title{Efficient inference for stochastic differential equation mixed-effects models using correlated particle  pseudo-marginal algorithms}

\author[add1]{Samuel Wiqvist\corref{cor1}}
\ead{samuel.wiqvist@matstat.lu.se}
\cortext[cor1]{Corresponding author}
\author[add2]{Andrew Golightly}
\author[add2]{Ashleigh T. McLean}
\author[add3]{Umberto Picchini}

\address[add1]{Centre for Mathematical Sciences, Lund University, Sweden}
\address[add2]{School of Mathematics, Statistics and Physics, Newcastle University, UK}
\address[add3]{Mathematical Sciences, Chalmers University of Technology and the University of Gothenburg, Sweden}

\begin{abstract}
Stochastic differential equation mixed-effects models (SDEMEMs) are flexible hierarchical models that are able to account for random variability inherent in the underlying time-dynamics, as well as the variability between experimental units and, optionally, account for measurement error. Fully Bayesian inference for state-space SDEMEMs is performed, using data at discrete times that may be incomplete and subject to measurement error. However, the inference problem is complicated by the typical intractability of the observed data likelihood which motivates the use of sampling-based approaches such as Markov chain Monte Carlo. A Gibbs sampler is proposed to target the marginal posterior of all parameter values of interest. The algorithm is made computationally efficient through careful use of blocking strategies and correlated pseudo-marginal Metropolis-Hastings steps within the Gibbs scheme. The resulting methodology is flexible and is able to deal with a large class of SDEMEMs. The methodology is demonstrated on three case studies, including tumor growth dynamics and neuronal data. The gains in terms of increased computational efficiency are model and data dependent, but unless bespoke sampling strategies requiring analytical derivations are possible for a given model, we generally observe an efficiency increase of one order of magnitude when using correlated particle methods together with our blocked-Gibbs strategy. 
\end{abstract}


\end{frontmatter}


\section{Introduction}
\label{sec:intro}

Stochastic differential equations (SDEs) are arguably the most used and studied stochastic dynamic models. SDEs allow the representation of stochastic time-dynamics, and are ubiquitous in applied research, most notably in finance \citep{steele2012stochastic}, systems biology \citep{Wilkinson06}, pharmacokinetic/pharmacodynamic modelling \citep{lavielle2014mixed} and neuronal modelling. SDEs extend the possibilities offered by ordinary differential equations (ODEs), by allowing random dynamics. As such, they can in principle replace ODEs in practical applications, to offer a richer mathematical representation for complex phenomena that are intrinsically non-deterministic.
However, in practice switching from ODEs to SDEs is usually far from trivial, due to the absence of closed form solutions to SDEs (except for the simplest toy problems), implying the need for numerical approximation procedures \citep{kloeden2013numerical}. Numerical approximation schemes, while useful for simulation purposes, considerably complicate statistical inference for model parameters. For reviews of inference strategies for SDE models, see e.g. \cite{Fuchs_2013} (including Bayesian approaches) and \cite{Sorensen04} (classical approaches). Generally, in the non-Bayesian framework, the literature for parametric inference approaches for SDEs is vast, however there is no inference procedure that is applicable to general nonlinear SDEs and that is also easy to implement on a computer. This is due to the lack of explicit transition densities for most SDE models. The problem is particularly difficult for measurements that are observed without error, i.e. Markovian observations. On the other hand, the Bayesian literature offers powerful solutions to the inference problem, when observations arise from state-space models. In our case, this means that if we assume that observations are observed with error, and that the latent process is a Markov process, then the literature based on sequential Monte Carlo (particle filters) is readily available in the form of pseudo-marginal methods \citep{andrieu09b}, and closely related particle MCMC methods \citep{andrieu10}, which we introduce in Section \ref{sec:pmmh}. 

Our goal is to produce novel Gibbs samplers embedding special types of pseudo-marginal algorithms allowing for exact Bayesian inference in a specific class of state-space SDE models. In this paper, we consider ``repeated measurement experiments'', modeled via mixed-effects, where the dynamics are Markov processes expressed via stochastic differential equations. These dynamics are assumed directly unobservable, i.e. are only observable up to measurement error. The practical goal is to fit observations pertaining to several ``units'' (i.e. independent experiments, such as measurements on different subjects) simultaneously, by formulating a state-space model having parameters randomly varying between the several individuals. The resulting model is typically referred to as a \textit{stochastic differential equation mixed-effects model} (SDEMEM). SDEMEMs are interesting because, in addition to explaining intrinsic stochasticity in the time-dynamics, they also take into account random variation between experimental units. The latter variation permits the understanding of between-subjects variability within a population.
When considered in a state-space model, these two types of variability (population variation and intrinsic stochasticity) are separated from the third source of variation, namely residual variation (measurement error). Thanks to their generality, and the ability to separate the three levels of variation, SDEMEMs have attracted attention, see e.g. \cite{donnet2013review} for a review and \cite{whitaker2016bayesian} for a more recent account. See also section \ref{sec:background} for a discussion on previous literature.


In the present work, we mainly focus on a general, \textit{plug-and-play} approach for exact Bayesian inference in SDEMEMs, meaning that analytic calculations are not necessary thanks to the flexibility of the underlying sequential Monte Carlo (SMC) algorithms. We also describe a non plug-and-play approach to handle specific situations. As in \cite{picchini17}, our random effects and measurement error can have arbitrary distributions, provided that the measurement error density can be evaluated point-wise. Unlike \cite{picchini17}, we use a Gibbs sampler to target the marginal parameter posterior. Subject specific, common and random effect population parameters are updated in separate blocks, with pseudo-marginal Metropolis-Hastings (PMMH) steps used to update the subject specific and common parameters, and Metropolis-Hastings (MH) steps used to update the random effect population parameters. 
We believe that, to date, our work results in the most general plug-and-play approach to inference for state-space SDEMEMs (a similar method has been concurrently and independently introduced (July 25 2019 on arXiv), in \cite{botha2019particle}; see the discussion in Section \ref{sec:disc}). However, the price to pay for such generality is that the use of pseudo-marginal methods guided by SMC algorithms is computationally consuming. In order to make pseudo-marginal methods scale better as the number of observations is increased, we exploit recent advances based on correlated PMMH (CPMMH). We combine CPMMH with a novel blocking strategy and show that it is possible to reduce considerably the number of required particles, and hence reduce the computational requirements for exact Bayesian inference. In our experiments, unless specific models admit bespoke efficient sampling strategies (e.g. Section \ref{sec:ou_neuronal} where it was possible to implement an advanced particle filter), CPMMH based algorithms with our novel blocking strategy are one order of magnitude more efficient than standard PMMH. Occasionally we even observed a 40-fold increase in efficiency, as in Section \ref{sec:ou}.

The remainder of this paper is organized as follows. Background literature is discussed in Section \ref{sec:background}. Stochastic differential mixed-effects models and the inference task are introduced in Section~\ref{sec:sdmem}. Our proposed approach to inference is described in Section~\ref{sec:pmmh}. Applications are considered in Section~\ref{sec:app}, including a simulation study considering an Ornstein-Uhlenbeck SDEMEM, a tumor-growth model and finally a challenging neuronal data case-study. A discussion is in Section~\ref{sec:disc}.
Julia and R codes can be found at \url{https://github.com/SamuelWiqvist/efficient_SDEMEM}. 

\section{Background literature}\label{sec:background}

Here we rapidly review key papers on inference for SDEMEMs, and refer the reader to \url{https://umbertopicchini.github.io/sdemem/} for a comprehensive list of publications. Early attempts at inference for SDEMEMs use methodology borrowed from standard (deterministic) nonlinear mixed-effects literature such as FOCE (first order conditional estimation) combined with the extended Kalman filter, as in \cite{overgaard2005non}. This approach can only deal with  SDEMEMs having a constant diffusion coefficient, see instead \cite{leander2015mixed} for an extension to state-dependent diffusion coefficients. The resulting inference in \cite{overgaard2005non} is approximate maximum likelihood estimation, and no uncertainty quantification is given. Moreover, only Gaussian random effects are allowed and measurement error is also assumed Gaussian. Other maximum likelihood approaches are in \cite{picchini2010stochastic} and \cite{picchini2011practical}, where a closed-form series expansion for the unknown transition density is found using the method in \cite{ait2008closed}, however the methodology can only be applied to reducible multivariate diffusions without measurement error. \cite{donnet2010bayesian} discuss inference for SDEMEMs in a Bayesian framework. They implement a Gibbs sampler when the SDE (for each subject) has an explicit solution, and consider Gaussian random effects and Gaussian measurement error. When no explicit solution exists, they approximate the diffusion process using the Euler-Maruyama approximation. The approach of \cite{donnet2013using} is of particular interest, since it is the first attempt to employ particle filters for inference in SDEMEMs: they construct an exact maximum likelihood strategy based on stochastic approximation EM (SAEM), where latent trajectories are ``proposed'' via particle Markov chain Monte Carlo. The major problem with using SAEM is the need for sufficient summary statistics for the ``complete likelihood'', which makes the methodology essentially impractical for arbitrarily complex models. \cite{delattre2013coupling} also use SAEM, but they avoid the need for the (usually unavailable) summary statistics for the complete likelihood, and propose trajectories using the extended Kalman filter instead of particle MCMC. Unlike in \cite{donnet2013using}, the inference in \cite{delattre2013coupling} is approximate and measurement error and random effects are required to be Gaussian. \cite{ruse2017multivariate} analyze multivariate diffusions under the conditions that the random effects are Gaussian distributed and that both fixed parameters and random effects enter linearly in the SDE. \cite{whitaker2017a} work with the Euler-Maruyama approximation and adopt a data augmentation approach to integrate over the uncertainty associated with the latent diffusion process, by employing carefully designed bridge constructs inside a Gibbs sampler. A linear noise approximation (LNA) is also considered. However, the limitations are that the observation equation has to be a linear combination of the latent states and measurement error has to be Gaussian. In addition, producing the bridge construct in the data augmentation approach or the LNA-based likelihood requires some careful analytic derivations. Consequently, neither approach can be regarded as a plug-and-play method (that is, a method that only requires forward simulation and evaluation of the measurement error density). In \cite{picchini17}, approximate and exact Bayesian approaches for a tumor growth study were considered: the approximate approach was based on synthetic likelihoods \citep{Wood10,price2018bayesian}, where summary statistics of the data are used for the inference, while exact inference used pseudo-marginal methodology via an auxiliary particle filter, which is suited to target measurements observed with a small error. It was found that using a particle approach to integrate out the random effects was very time consuming. Even though the data set was small (comprising 5-8 subjects to fit, depending on the experimental group, and around 10 observations per subject), the number of particles required to approximate each individual likelihood was in the order of thousands. This is very time consuming when the number of ``subjects'' (denoted $M$ in the rest of this work) increases.

\section{Stochastic differential mixed-effects models}
\label{sec:sdmem}

Consider the case where we have $M$ experimental units randomly chosen
from a theoretical population. Our goal is to perform inference based on simultaneously fitting all data from the $M$ units. Now assume that the experiment we are analyzing consists in observing a stochastically evolving dynamic process,
and that associated with each unit $i$ is a
continuous-time $d$-dimensional It\^o process $\{X_t^i, t\geq 0\}$
governed by the SDE
\begin{equation}
dX_t^i=\alpha(X_t^i,\kappa,\phi^i,D^i)\,dt+\sqrt{\beta(X_t^i,\kappa,\phi^i,D^i)}\,dW_t^i,
\quad X_0^i=x_0^i, \quad i=1,\ldots,M. \label{eqn:sdmem}
\end{equation}
Here, $\alpha$ is a $d$-vector of drift functions, the 
diffusion coefficient $\beta$ is a $d \times d$ positive 
definite matrix with a square root representation 
$\sqrt{\beta}$ such that $\sqrt{\beta}\sqrt{\beta}^T=\beta$, $W_t^i$ is a $d$-vector of (uncorrelated) standard 
Brownian motion processes and $D^i$ are unit-specific static or time-dependent deterministic input (e.g. covariates, forcing functions), see e.g. \cite{leander2015mixed}. The $p$-vector parameter 
$\kappa=(\kappa_{1},\ldots,\kappa_{p})^T$ is common 
to all units whereas the $q$-vectors  $\phi^i=(\phi_{1}^{i},\ldots,\phi_{q}^{i})^T$, 
$i=~1,\ldots,M$, are unit-specific random effects. 
In the most general random effects 
scenario we let $\pi(\phi^i|\eta)$ denote the joint distribution 
of $\phi^i$, parameterised by the $r$-vector $\eta=(\eta_{1},\ldots,\eta_{r})^T$. 
The model defined by (\ref{eqn:sdmem}) allows for differences between 
experimental units through different realizations of the Brownian motion paths $W_t^i$ 
and the random effects $\phi^i$, accounting for inherent stochasticity within
a unit, and variation between experimental units respectively. 

We assume that each experimental unit $\{X_t^i,t\geq 0\}$ cannot be
observed exactly, but observations
$y^i=(y_{1}^i,\ldots,y_{n}^i)^T$ are available. Without loss of generality, we assume units are observed at the same integer time points $\{1,2,...,n\}$, that is in the following we write $n$ instead of, say, $n_i$ for all $i$. However this is only for convenience of notation, and we could easily accommodate the possibility of different units $i$ having different values $n_i$ and that, in turn, units are observed at different sets of times. The observations are assumed conditionally independent (given the latent process) and we link
them to the latent process via 
\begin{equation}\label{eqn:obs}
    Y_t^i = h(X_t^i,S^i,\epsilon_t^i), \qquad \epsilon_t^i|\xi\indep p_\epsilon(\xi), \qquad i=1,...,M
\end{equation}
where $Y_t^i$ is a $d_o$-vector, $\epsilon_t$ is a random $d_o$-vector, $d_o\leq d$, $\epsilon_t^i$ is the measurement noise, $S^i$ is (as $D^i$) a unit-specific deterministic input, and $h(\cdot)$ is a possibly nonlinear function of its arguments. In the applications in Section \ref{sec:app} we have $D^i=S^i=\emptyset$, the empty set, for every $i$, and hence for simplicity of notation we disregard $D^i$ and $S^i$ in the rest of the paper. However having non-empty sets does not introduce any additional complication to our methodology. Notice, the possibility to have $d_0<d$ implies that we may have some coordinate of the $\{X_t^i\}$ system that is unobserved at some (or all) $t$.
We denote the density linking $Y_t^i$ and $X_t^i$ by $\pi(y_t^i|x_t^i,\xi)$. 
An important special case that arises from our flexible observation model is when $h(X_t^i,\epsilon_t^i)=F^T X_t^i + \epsilon_t^i$ for a constant matrix $F$ and $\epsilon_t^i|\Sigma\indep N(0,\Sigma)$, allowing for observation of a linear combination of components of $X_t^i$, subject to additive Gaussian noise. Notice that our methodology in Sections \ref{sec:bayes}--\ref{sec:tuning} can be applied to an arbitrary $h(\cdot)$, provided this can be evaluated pointwise for any value of its arguments. For example, in Section \ref{sec:tum} we have that $h(\cdot)$ is the logarithm of the sum of the components of a bivariate $X_t^i$.

We refer to the model constituted by the system \eqref{eqn:sdmem}-\eqref{eqn:obs} as a SDEMEM. This is a state-space model, due to the Markov property of the It\^o processes $\{X_t^i, t\geq 0\}$, and the assumption of conditional independence of observations on latent processes. The model is flexible: equation \eqref{eqn:sdmem} explains the intrinsic stochasticity in the dynamics (via $\beta$) and the variation between-units (via the random effects $\phi^i$), while \eqref{eqn:obs} explains residual variation (measurement error, via $\xi$).

\subsection{Bayesian inference}
\label{sec:bayes}

Denote with $x=(x^1,\ldots,x^M)^T$ the set of unobserved states collected across all $M$ diffusion processes $\{X_t^i\}$ at the same set of integer times $\{1,2,...,n\}$ as for data $y=(y^1,\ldots,y^M)^T$. 
Then given data $y=(y^1,\ldots,y^M)^T$, latent values $x$, the joint posterior for the common parameters $\kappa$, fixed/random
effects $\phi=(\phi^1,\ldots,\phi^M)^T$, hyperparameters $\eta$ and measurement error parameters 
$\xi$ is 
\begin{equation}
\pi(\kappa,\eta,\xi,\phi,x|y)\propto
\pi(\kappa,\eta,\xi)\pi(\phi|\eta)\pi(x|\kappa,\phi)\pi(y|x,\xi) \label{eqn:posterior}
\end{equation}
where $\pi(\kappa,\eta,\xi)$  is the joint prior density ascribed to $\kappa$, 
$\eta$ and $\xi$. These three parameters may be assumed \emph{a priori} independent, and then we can write $\pi(\kappa,\eta,\xi)=\pi(\kappa)\pi(\eta)\pi(\xi)$, though this needs not be the case and we can easily assume \emph{a priori} correlated parameters. In addition we have that
\begin{equation}\label{eqn:phi}
\pi(\phi|\eta)=\prod_{i=1}^{M}\pi(\phi^i|\eta),
\end{equation}
\begin{equation}\label{eqn:y}
\pi(y|x,\xi)= \prod\limits_{i=1}^{M}\prod\limits_{j=1}^{n}\pi(y_{j}^i|x_{j}^i,\xi)
\end{equation}
and
\begin{equation}\label{eqn:xdens}
\pi(x|\kappa,\phi)=\prod\limits_{i=1}^{M}\pi(x_1^i)\prod\limits_{j=2}^{n}\pi(x^i_{j}|x^i_{j-1},\kappa,\phi^i).
\end{equation}
Note that $\pi(x^i_{j}|x^i_{j-1},\kappa,\phi^i)$ will be typically intractable. In this case, we assume 
that it is possible to generate draws (up to arbitrary accuracy) from $\pi(x^i_{j}|x^i_{j-1},\kappa,\phi^i)$ 
using a suitable numerical approximation. For example, the Euler-Maruyama approximation of (\ref{eqn:sdmem}) is
\[
\Delta X_t^i\equiv X^i_{t+\Delta t}-X^i_t=\alpha(X_t^i,\kappa,\phi^i)\,\Delta t+\sqrt{\beta(X_t^i,\kappa,\phi^i)}\,\Delta W_t^i
\]
and therefore
\begin{equation}\label{em}
X^i_{t+\Delta t}=X^i_t+\alpha(X_t^i,\kappa,\phi^i)\,\Delta t+\sqrt{\beta(X_t^i,\kappa,\phi^i)}\,\Delta W_t^i
\end{equation}
where $\Delta W_t^i\sim N(0,I_d\Delta t)$ and the time-step $\Delta t$, which need not be the inter-observation time, 
is chosen by the practitioner to balance accuracy and efficiency.

In what follows, we assume that interest lies in the marginal
posterior for all parameters, given by $\pi(\kappa,\eta,\xi,\phi | y)=\int \pi(\kappa,\eta,\xi,\phi,x | y)dx$, where
\begin{align}\label{target}
\pi(\kappa,\eta,\xi,\phi | y)&\propto\pi(\kappa)\pi(\eta)\pi(\xi)\pi(\phi|\eta)\pi(y|\kappa,\xi,\phi)\\
&\propto  \pi(\kappa)\pi(\eta)\pi(\xi)\prod\limits_{i=1}^M\pi(\phi^i|\eta)\pi(y^i|\kappa,\xi,\phi^i). 
\end{align}
This factorization suggests a Gibbs sampler with separate blocks for each parameter vector that
sequentially takes draws from the full conditionals 
\begin{enumerate}
\item $\pi(\phi|\kappa,\eta,\xi,y)\propto \prod_{i=1}^M \pi(\phi^i|\eta)\pi(y^i|\kappa,\xi,\phi^i)$,
\item $\pi(\kappa|\eta,\xi,\phi,y)=\pi(\kappa|\phi,\xi,y)\propto \pi(\kappa) \prod_{i=1}^M \pi(y^i|\kappa,\xi,\phi^i)$, 
\item $\pi(\xi|\kappa,\eta,\phi,y)=\pi(\xi|\kappa,\phi,y)\propto \pi(\xi)\prod_{i=1}^M \pi(y^i|\kappa,\xi,\phi^i) $,
\item $\pi(\eta|\kappa,\xi,\phi,y)=\pi(\eta|\phi)\propto \pi(\eta)\prod_{i=1}^M \pi(\phi^i|\eta)$. 
\end{enumerate} 
Of course, in practice, the observed individual data likelihood $\pi(y^i|\kappa,\xi,\phi^i)=\int p(y^i,x^i|\kappa,\xi,\phi^i)dx^i$ will be intractable. 
In what follows, therefore, we consider a Metropolis-within-Gibbs strategy, and in particular introduce auxiliary 
variables $u$ to allow pseudo-marginal Metropolis-Hastings updates.

\section{A pseudo-marginal approach}
\label{sec:pmmh}
Consider again the intractable target in (\ref{target}) and suppose that we can unbiasedly estimate the intractable 
observed data likelihood $\pi(y|\kappa,\xi,\phi)=\int p(y,x|\kappa,\xi,\phi)dx$. To this end let 
\[
\hat{\pi}_u(y|\kappa,\xi,\phi)=\prod_{i=1}^M \hat{\pi}_{u^i}(y^i|\kappa,\xi,\phi^i)
\]
denote a (non-negative) unbiased estimator of $\pi(y|\kappa,\xi,\phi)$, where $u=(u^1,\ldots,u^M)^T$ is the collection 
of auxiliary (vector) variables used to produce the corresponding estimate, with density $\pi(u)=\prod_{i=1}^M g(u^i)$. In the context of inference for SDEs, the $u$ may be the collection of pseudo-random standard Gaussian draws, these being necessary to simulate increments of the Brownian motion paths when implementing a numerical scheme such as Euler-Maruyama (Section \ref{sec:like}), or produce draws from transition densities (in the rare instances when these are known). Notice in fact that the $u$ need not have a specific distribution, though in stochastic simulation we need access to pseudo-random variates that are often uniform or Gaussian distributed \citep{devroye}. When inference methods use particle filters, pseudo-random variates are also employed in the resampling step, and hence these variates can be included into $u$.

Now, the pseudo-marginal Metropolis-Hastings (PMMH) scheme targets 
\begin{align}\label{target2}
\pi(\kappa,\eta,\xi,\phi,u | y)&\propto \pi(\kappa)\pi(\eta)\pi(\xi)\pi(\phi|\eta)\hat{\pi}_{u}(y|\kappa,\xi,\phi)\pi(u)
\end{align}
for which it is easily checked that 
\begin{align*}
\int\pi(\kappa,\eta,\xi,\phi,u | y)du &\propto   \pi(\kappa)\pi(\eta)\pi(\xi)\pi(\phi|\eta)\int \hat{\pi}_{u}(y|\kappa,\xi,\phi)\pi(u)du\\
&\propto \pi(\kappa,\eta,\xi,\phi | y).
\end{align*}
Hence, marginalising out $u$ gives the marginal parameter posterior in (\ref{target}). Directly targeting the high dimensional posterior $\pi(\kappa,\eta,\xi,\phi,u | y)$ with PMMH is likely to give very small acceptance rates. The structure of the SDMEM naturally admits a Gibbs sampling strategy. We outline our novel Gibbs samplers in the next section. 

\subsection{Gibbs sampling and blocking strategies}
\label{sec:gibbs}
The form of (\ref{target2}) immediately suggests a Gibbs sampler 
that
sequentially takes draws from the full conditionals. However, we can design two types of Gibbs samplers. Our first, novel strategy is denoted ``naive Gibbs'', where the $u^i$ are updated with both the subject specific and common parameters.\\

\noindent
\textbf{Naive Gibbs:}
\begin{enumerate}
\item $\pi(\phi^i,u^i|\kappa,\eta,\xi,y^i)\propto \pi(\phi^i|\eta)\hat{\pi}_{u^i}(y^i|\kappa,\xi,\phi^i)g(u^i)$, $i=1,\ldots,M$,
\item $\pi(\kappa,u|\eta,\xi,\phi,y,u)=\pi(\kappa,u|\phi,\xi,y)\propto \pi(\kappa) \prod_{i=1}^M \hat{\pi}_{u^i}(y^i|\kappa,\xi,\phi^i)g(u^i)$, 
\item $\pi(\xi,u|\kappa,\eta,\phi,y,u)=\pi(\xi,u|\kappa,\phi,y)\propto \pi(\xi)\prod_{i=1}^M \hat{\pi}_{u^i}(y^i|\kappa,\xi,\phi^i)g(u^i)$,
\item $\pi(\eta|\kappa,\xi,\phi,y,u)=\pi(\eta|\phi)\propto \pi(\eta)\prod_{i=1}^M \pi(\phi^i|\eta)$. 
\end{enumerate}
Note that step 1 consists of a set of draws of $M$ conditionally independent random variables since
\begin{align*}
\pi(\phi,u|\kappa,\eta,\xi,y)&=\prod_{i=1}^M \pi(\phi^i,u^i|\kappa,\eta,\xi,y^i).
\end{align*}
Hence, step 1 gives a sample from $\pi(\phi,u|\kappa,\eta,\xi,y)$. Draws from the full conditionals in 1-3 can be obtained by using Metropolis-Hastings within Gibbs. Taking the $[\phi^i,u^i]$ block as an example,  
we use a proposal density of the form $q(\phi^{i*}|\phi^i)g(u^{i*})$ and accept a move from $[\phi^i,u^i]$ 
to $[\phi^{i*},u^{i*}]$ with probability
\[
\min\left\{1\,,\, \frac{\pi(\phi^{i*}|\cdot)}{\pi(\phi^{i}|\cdot)}\times 
\frac{\hat{\pi}_{u^{i*}}(y^i|\phi^{i*},\cdot)}{\hat{\pi}_{u^{i}}(y^i|\phi^{i},\cdot)}\times 
\frac{q(\phi^{i}|\phi^{i*})}{q(\phi^{i*}|\phi^{i})}\right\}.
\]
Effectively, samples from the full conditionals in 1-3 are obtained via draws from pseudo-marginal MH kernels. 

The above strategy is somewhat naive, since the auxiliary variables 
$u$ need only be updated once per Gibbs iteration, instead in steps 1 to 3 of the naive Gibbs procedure vectors $u^i$ are simulated anew in each of the three steps (notice $g(u^i)$ appears in each of the first three steps). We therefore propose to update the blocks $[\phi^i,u^i]$, $i=1,\ldots,M$ in step 1 only, and condition on the most recent value of $u$ in the remaining steps. We call this second, novel strategy ``blocked Gibbs''. \\

\noindent
\textbf{Blocked Gibbs:}
\begin{enumerate}
\item $\pi(\phi^i,u^i|\kappa,\eta,\xi,y^i)\propto  \pi(\phi^i|\eta)\hat{\pi}_{u^i}(y^i|\kappa,\xi,\phi^i)g(u^i)$, $i=1,\ldots,M$,
\item $\pi(\kappa|\eta,\xi,\phi,y,u)=\pi(\kappa|\phi,\xi,y,u)\propto \pi(\kappa) \prod_{i=1}^M \hat{\pi}_{u^i}(y^i|\kappa,\xi,\phi^i)$, 
\item $\pi(\xi|\kappa,\eta,\phi,y,u)=\pi(\xi|\kappa,\phi,y,u)\propto \pi(\xi)\prod_{i=1}^M \hat{\pi}_{u^i}(y^i|\kappa,\xi,\phi^i) $,
\item $\pi(\eta|\kappa,\xi,\phi,y,u)=\pi(\eta|\phi)\propto \pi(\eta)\prod_{i=1}^M \pi(\phi^i|\eta)$. 
\end{enumerate}
The aim of blocking in this way is to reduce the variance of the acceptance probability associated with steps 2 and 3, which involve the product of $M$ estimates as opposed to a single estimate in each constituent part of step 1. Also, notice $g(u^i)$ appears only in the first step. The effect of blocking in this way is explored empirically in Section~\ref{sec:app}. 

\subsection{Estimating the likelihood}
\label{sec:like}
It remains that we can generate  non-negative unbiased estimates $\hat{\pi}_u(y|\kappa,\xi,\phi)$. This can be achieved 
by running a sequential Monte Carlo procedure, also known as particle filter. The simplest approach is to use the bootstrap particle filter \citep{stewart1992use,gordon93} (see also \cite{Kunsch2013}) that, for a single experimental unit, 
recursively draws from the filtering distribution $\pi(x_t^i|y_{1:t}^i,\kappa,\xi,\phi^i)$ for each $t=1,\ldots,n$. Here,  $y_{1:t}^i$ denotes the  observations of experiment $i$ for time-steps $1,\dots,t$. Essentially, 
a sequence of importance sampling and resampling steps are used to propagate a weighted sample $\{(x_{t,k}^i,w(u_{t,k}^i)),k=1,\ldots,N_{i}\}$ 
from the filtering distribution, where $N_i$ is the number of particles for unit $i$. Note that we let the weight depend explicitly on the $t$-th component of the auxiliary variable 
$u^i=(u_1^i,\ldots,u_n^i)$, associated with experimental unit $i$. At time $t$, the particle filter uses the approximation
\begin{equation}\label{pftarget}
\hat{\pi}(x_{t}^i|y_{1:t}^i,\kappa,\xi,\phi^i)\propto \pi(y_{t}^i|x_{t}^i,\xi)\sum_{k=1}^{N_{i}}\pi(x_{t}^i|x_{t-1,k}^{i},\kappa,\phi^i)w(u_{t-1,k}^i).
\end{equation}
A simple importance sampling/resampling strategy follows, where particles are resampled (with replacement) in proportion to their weights, 
propagated via $x_{t,k}^i=f_{t}(u_{t,k}^i)\sim \pi(\cdot|x_{t-1,k}^{i},\kappa,\phi^i)$ and reweighted by $p(y_{t}^i|x_{t,k}^i,\xi)$. Here, 
$f_t(\cdot)$ is a deterministic function of $u_{t,k}^i$ (as well as the parameters and previous latent state, suppressed for simplicity) that gives an explicit connection between 
the particles and auxiliary variables. An example of $f_t(\cdot)$ is to take the Euler-Maruyama approximation
\[
f_{t}(u_{t,k}^i)=x^i_{t-1,k}+\alpha(x_{t-1,k}^i,\kappa,\phi^i)\,\Delta t+\sqrt{\beta(x_{t-1,k}^i,\kappa,\phi^i)\Delta t}\times u_{t,k}^i
\]    
where $u_{t,k}^i\sim N(0,I_d)$ and $\Delta t$ is a suitably chosen time-step. In practice, unless $\Delta t$ is sufficiently small to allow 
an accurate Euler-Maruyama approximation, $f_{t}(u_{t,k}^i)$ will describe recursive application of the numerical approximation.  

\begin{algorithm}[ht]
\footnotesize
\caption{Bootstrap particle filter for experimental unit $i$}\label{BPF}
\textbf{Input:} parameters $\kappa$, $\phi^i$, $\xi$, auxiliary variables $u^i$, data $y^i$ and the number of particles $N_{i}$.\\
\textbf{Output:} estimate $\hat{\pi}_{u^i}(y^i|\kappa,\xi,\phi^i)$ of the observed data likelihood.
\begin{enumerate}
\item Initialisation ($t=1$). 
\begin{itemize}
\item[(a)] \textbf{Sample} the prior. Put $x_{1,k}^{i}=f_1(u_{1,k}^i)\sim \pi(\cdot)$, $k=1,\ldots,N_{i}$.
\item[(b)] \textbf{Compute} the weights. For $k=1,\ldots,N_{i}$ set
\[
\tilde{w}(u_{1,k}^i)=\pi(y_{1}^i|x_{1,k}^{i},\xi), \qquad w(u_{1,k}^{i})=\frac{\tilde{w}(u_{1,k}^i)}{\sum_{j=1}^{N_{i}}\tilde{w}(u_{1,j}^i)}.
\]
\item[(c)] \textbf{Update} observed data likelihood estimate. Compute $\hat{\pi}_{u_1^i}(y_{1}^i|\kappa,\xi,\phi^i)=\sum_{k=1}^{N_{i}}\tilde{w}(u_{1,k}^i)/N_{i}$.
\end{itemize}
\item For times $t=2,3,\ldots ,n$:
\begin{itemize}
\item[(b')] \textbf{(optional) Sorting.} Use Euclidean sorting on particles  $\{x_{t-1,1}^i,...,x_{t-1,N_i}^i\}$ if using CPMMH.
\item[(b)] \textbf{Resample.} Obtain ancestor indices $a_{t-1}^k$, $k=1,\ldots,N_{i}$ using systematic resampling on the collection of weights 
$\{w(u_{t-1,1}^i),\ldots,w(u_{t-1,N_{i}}^i)\}$.
\item[(c)] \textbf{Propagate.} Put $x_{t,k}^{i}=f_t({u}_{t,k}^i)\sim \pi\big(\cdot|x_{t-1,a_{t-1}^k}^i,\kappa,\xi,\phi^i\big)$, $k=1,\ldots,N_{i}$. 
\item[(d)] \textbf{Compute} the weights. For $k=1,\ldots,N_{i}$ set
\[
\tilde{w}(u_{t,k}^i)=\pi(y_{t}^i|x_{t,k}^{i},\xi), \qquad w(u_{t,k}^{i})=\frac{\tilde{w}(u_{t,k}^i)}{\sum_{j=1}^{N_{i}}\tilde{w}(u_{t,j}^i)}.
\]
\item[(e)] \textbf{Update} observed data likelihood estimate. Compute
\[
\hat{\pi}_{u_{1:t}^i}(y_{1:t}^i|\kappa,\xi,\phi^i)=\hat{\pi}_{u_{1:t-1}^i}(y_{1:t-1}^i|\kappa,\xi,\phi^i)\hat{\pi}_{u_{t}^i}(y_{t}^i|y_{1:t-1}^i,\kappa,\xi,\phi^i)
\]
where $\hat{\pi}_{u_{t}^i}(y_{t}^i|y_{1:t-1}^i,\kappa,\xi,\phi^i)=\sum_{k=1}^{N_{i}}\tilde{w}(u_{t,k}^i)/N_{i}$.
\end{itemize}
\end{enumerate}
\end{algorithm}   

Algorithm~\ref{BPF} provides a complete description of the bootstrap particle filter when applied to a single experimental unit. However notice the addition of a non-standard and optional sorting step 2b', which turns useful when implementing a correlated pseudo-marginal approach, as described in Section \ref{sec:corps}. For the resampling step we follow \cite{deligiannidis2016} among others 
and use systematic resampling (see e.g. \cite{murray2016}), which only requires 
simulating a single uniform random variable at each time point. It is straightforward to augment the auxiliary variable $u^i$ to include the 
random variables used in the resampling step. As a by-product of the particle filter, 
the observed data likelihood $\pi(y^i|\kappa,\xi,\phi^i)$ can be estimated 
via the quantity 
\begin{equation}
\hat{\pi}_{u^i}(y^i|\kappa,\xi,\phi^i) =N_{i}^{-n}\prod_{t=1}^{n}\sum_{k=1}^{N_{i}}\tilde{w}(u_{t,k}^i).
\end{equation}
Moreover, the corresponding estimator can be shown to be unbiased \citep{delmoral04,pitt12}.

The full Gibbs sampler for generating draws from the joint posterior (\ref{target2}) is given by Algorithm~\ref{algGibbs}. 
For ease of exposition, we have blocked the updates for $\kappa$ and $\xi$, but note that the use of separate updates for these parameters 
is straightforward. The precise implementation of step 4 of the Gibbs sampler is likely to be example specific, and we anticipate that a direct draw 
of $\eta^{(j)}\sim \pi(\cdot|\phi^{(j)})$ will often be possible. For example when the components of $\phi$ are assumed to be normally 
distributed and $\eta$ consists of the corresponding means and precisions, for which a semi-conjugate prior specification is possible, see Section \ref{sec:ou}.     

\begin{algorithm}[ht]
\footnotesize
\caption{Blocked Gibbs sampler}\label{algGibbs}
\textbf{Input:} Data $y$, initial parameter values $\phi$, $\kappa$, $\xi$, $\eta$ and number of iterations $n_{\textrm{iters}}$.\\
\textbf{Output:} $\{\phi^{(j)},\kappa^{(j)},\xi^{(j)},\eta^{(j)}\}_{j=1}^{n_{\textrm{iters}}}$.
\begin{enumerate}
\item Initialise $\phi^{(0)}=(\phi^{1,(0)},\ldots,\phi^{M,(0)})$, $\kappa^{(0)}$, $\xi^{(0)}$. Draw $u^{i,(0)}\sim g(\cdot)$ and run Algorithm~\ref{BPF} for $i=1,\ldots,M$ 
with $u^{i,(0)}$, $\phi^{i,(0)}, \kappa^{(0)}, \xi^{(0)}$ and $y^i$ to obtain $\hat{\pi}_{u^{i,(0)}}(y^i|\kappa^{(0)},\xi^{(0)},\phi^{i,(0)})$. Set the iteration counter $j=1$.
\item Update subject specific parameters. For $i=1,\ldots,M$: 
\begin{itemize}
\item[(a)] Propose $u^{i*}\sim g(\cdot)$ and $\phi^{i*}\sim q(\cdot|\phi^{i,(j-1)})$.
\item[(b)] Compute $\hat{\pi}_{u^{i*}}(y^i|\kappa^{(j-1)},\xi^{(j-1)},\phi^{i*})$ by running Algorithm~\ref{BPF} with 
$u^{i*}$, $\phi^{i*}$, $\kappa^{(j-1)}$, $\xi^{(j-1)}$ and $y^i$.
\item[(c)] With probability
\begin{equation}\label{aprob1}
\min\left\{1\,,\, \frac{\pi(\phi^{i*}|\eta)}{\pi(\phi^{i,(j-1)}|\eta)}\times 
\frac{\hat{\pi}_{u^{i*}}(y^i|\kappa^{(j-1)},\xi^{(j-1)},\phi^{i*})}{\hat{\pi}_{u^{i,(j-1)}}(y^i|\kappa^{(j-1)},\xi^{(j-1)},\phi^{i,(j-1)})}\times 
\frac{q(\phi^{i,(j-1)}|\phi^{i*})}{q(\phi^{i*}|\phi^{i,(j-1)})}\right\}
\end{equation}
put $\phi^{i,(j)}=\phi^{i*}$ and $u^{i,(j)}=u^{i*}$. Otherwise, store the current values $\phi^{i,(j)}=\phi^{i,(j-1)}$ and $u^{i,(j)}=u^{i,(j-1)}$.
\end{itemize}
\item Update common parameters.
\begin{itemize}
\item[(a)] Propose $(\kappa^*,\xi^*)\sim q(\cdot|\kappa^{(j-1)},\xi^{(j-1)})$.  
\item[(b)] Compute $\hat{\pi}_{u^{(j)}}(y|\kappa^*,\xi^*,\phi^{(j)})=\prod_{i=1}^{M}\hat{\pi}_{u^{i,(j)}}(y^i|\kappa^{*},\xi^{*},\phi^{i,(j)})$ by running Algorithm~\ref{BPF} 
for $i=1,\ldots,M$ with $u^{i,(j)}$, $\phi^{i,(j)}$, $\kappa^{*}$, $\xi^{*}$ and $y^i$.
\item[(c)] With probability
\begin{equation}\label{aprob2}
\min\left\{1\,,\, \frac{\pi(\kappa^*)\pi(\xi^*)}{\pi(\kappa^{(j-1)})\pi(\xi^{(j-1)})}\times 
\frac{\hat{\pi}_{u^{(j)}}(y|\kappa^{*},\xi^{*},\phi^{(j)})}{\hat{\pi}_{u^{(j)}}(y|\kappa^{(j-1)},\xi^{(j-1)},\phi^{(j)})}\times
\frac{q(\kappa^{(j-1)},\xi^{(j-1)}|\kappa^*,\xi^*)}{q(\kappa^*,\xi^*|\kappa^{(j-1)},\xi^{(j-1)})} \right\}
\end{equation}
put $(\kappa^{(j)},\xi^{(j)})=(\kappa^{*},\xi^*)$. Otherwise, store the current values $(\kappa^{(j)},\xi^{(j)})=(\kappa^{(j-1)},\xi^{(j-1)})$.
\end{itemize}
\item Update random effect population parameters. Draw $\eta^{(j)}\sim \pi(\cdot|\phi^{(j)})$.
\item If $j=n_{\textrm{iters}}$, stop. Otherwise, set $j:=j+1$ and go to step 2.
\end{enumerate}
\end{algorithm}

Executing Algorithm~\ref{algGibbs} requires $n\sum_{i=1}^{M} N_i$ draws from the transition density governing the SDE in (\ref{eqn:sdmem}) 
\emph{per iteration}. In scenarios where the transition density is intractable, draws of a suitable 
numerical approximation are required. For example, we may use the Euler-Maruyama discretisation with time step $\Delta t=1/m$, where $m\geq 1$ is chosen to limit the associated discretisation bias (and typically $m\gg 1$). In this case, order $m n \sum_{i=1}^{M} N_i$ draws of (\ref{em}) 
are required. As discussed by \cite{andrieu10}, 
the number of particles per experimental unit, $N_i$, should be scaled in proportion to the number of data points $n$. Consequently, the use of PMMH kernels is 
likely to be computationally prohibitive in practice. We therefore consider the adaptation of a recently proposed correlated PMMH 
method for our problem.

\subsection{A correlated pseudo-marginal approach}
\label{sec:corps}

Consider again the task of sampling the full conditional $\pi(\phi^i,u^i|\kappa,\eta,\xi,y^i)$ associated with the $i$th 
experimental unit. In steps 2(a--c) of Algorithm~\ref{algGibbs}, a (pseudo-marginal) Metropolis-Hastings step is used whereby the auxiliary variables 
$u^i$ are proposed from the associated pdf $g(\cdot)$ (notice we could introduce a subject-specific $g_i(\cdot)$, but we refrain from doing so in the interest of a lighter notation). As discussed by \cite{deligiannidis2016} 
(see also \cite{dahlin2015}), the proposal kernel need not be restricted to the use of $g(u^i)$. The correlated PMMH (CPMMH) scheme 
generalises the PMMH scheme by generating a new $u^{i*}$ from $K(u^{i*}|u^i)$ where $K(\cdot|\cdot)$ satisfies the detailed balance equation
\begin{equation}\label{Kdb}
g(u^i)K(u^{i*}|u^i)=g(u^{i*})K(u^i|u^{i*}).
\end{equation}
It is then straightforward to show that a MH scheme with proposal kernel $q(\phi^{i*}|\phi^{i}) K(u^{i*}|u^i)$ and acceptance probability (\ref{aprob1}) 
satisfies detailed balance with respect to the target $\pi(\phi^i,u^i|\kappa,\eta,\xi,y^i)$.

We take $g(u^i)$ as a standard Gaussian density and $K(u^{i*}|u^i)$ as the kernel associated with a Crank--Nicolson proposal \citep{deligiannidis2016}. Hence
\[
g(u^i)=\textrm{N}\left(u^i;\,0\,,\,I_d\right)\qquad \textrm{and} \qquad K(u^{i*}|u^i)=\textrm{N}\left(u^{i*};\,\rho u^i\,,\,\left(1-\rho^2\right)I_d\right)
\]
where $I_d$ is the identity matrix whose dimension $d$ is determined by the number of elements in $u^i$. The parameter $\rho$ is chosen 
to be close to 1, to induce strong positive correlation between $\hat{\pi}_{u^{i}}(y^i|\kappa,\Sigma,\phi^{i})$ and 
$\hat{\pi}_{u^{i*}}(y^i|\kappa,\Sigma,\phi^{i*})$, thus reducing the variance of the acceptance probability in (\ref{aprob1}), which is beneficial because it reduces the chance of accepting an overestimation of the likelihood function. Taking $\rho=0$ 
gives the special case that $K(u^{i*}|u^i)=g(u^{i*})$, which corresponds to the standard PMMH. Iteration $j$ of step 2 of Algorithm~\ref{algGibbs} then becomes
\begin{enumerate}
\item[2.] For $i=1,\ldots,M$:
\begin{itemize}
\item[(a)] Propose $\phi^{i*}\sim q(\cdot|\phi^{i,(j-1)})$. Draw $\omega\sim \textrm{N}(0,I_d)$ and put $u^{i*}=\rho u^{i,(j-1)}+\sqrt{1-\rho^2}\omega$. 
\item[(b)] Compute $\hat{\pi}_{u^{i*}}(y^i|\kappa^{(j-1)},\xi^{(j-1)},\phi^{i*})$ by running Algorithm~\ref{BPF} with 
$u^{i*}$, $\phi^{i*}$, $\kappa^{(j-1)}$, $\xi^{(j-1)}$ and $y^i$.
\item[(c)] With probability given by (\ref{aprob1}) 
put $\phi^{i,(j)}=\phi^{i*}$ and $u^{i,(j)}=u^{i*}$. Otherwise, store the current values $\phi^{i,(j)}=\phi^{i,(j-1)}$ and $u^{i,(j)}=u^{i,(j-1)}$.
\end{itemize}
\end{enumerate}
Care must be taken here when executing Algorithm~\ref{BPF} in Step 2(b). Upon changing $\phi^i$ and $u^i$, 
the effect of the resampling step is likely to prune out different particles, thus breaking the 
correlation between successive estimates of observed data likelihood. Sorting the particles before resampling can alleviate this problem \citep{deligiannidis2016}. We follow \cite{choppala2016} (see also \cite{golightly2019}) by using a simple Euclidean sorting procedure which, for the case of a 1-dimensional latent state (e.g. when $\dim(X_t^i)=1$ for every $t$) implies, prior to resampling the particles, to sort the particles from the smallest to the largest. This is step 2b' in algorithm \ref{BPF}, denoted ``optional'' as it only applies to CPMMH, not PMMH.

\subsection{Tuning the number of particles for likelihood approximation}
\label{sec:tuning}

It remains that we can choose the number of particles $N_i$ to be used to obtain 
estimates of the observed data likelihood contributions 
$\hat{\pi}_{u^i}(y^i|\kappa,\xi,\phi^i)$. Note that we allow a different number of particles per experimental unit to accommodate differing lengths of the $y^i$ and potential model misspecification at the level of an individual unit. In the case of PMMH, a simple strategy is to fix $\phi^i$, 
$\kappa$ and $\xi$ at some central posterior value (obtained from a pilot run), 
and choose $N_i$ so that the variance of the log-likelihood 
(denoted $\sigma^{2}_{N_{i}}$) is around 2 \citep{doucet15,sherlock2015}. When using a CPMMH kernel, we follow \cite{tran2016,choppala2016} by 
choosing $N_{i}$ so that $\sigma^{2}_{N_i}=2.16^2/(1-\rho_{l}^2)$ where 
$\rho_l$ is the estimated correlation between $\log \hat{\pi}_{u^i}(y^i|\kappa,\xi,\phi^i)$ and $\log \hat{\pi}_{u^{i*}}(y^i|\kappa,\xi,\phi^i)$. Hence, an initial pilot run (with the number of particles set at some conservative value) is required to determine plausible values of the parameters. This pilot run can also be used to give estimates of $var(\phi^i|y^i)$, $i=1,\ldots,M$, each of which can subsequently be used as the innovation variance in a Gaussian random walk proposal for $\phi^i$.   

\subsection{Tuning the proposal distributions} \label{sec:tune_prop_dist}
The block structure of the Gibbs sampler (Algorithm~\ref{algGibbs}) requires two proposal densities:  $\phi^{i*}\sim q(\cdot|\phi^{i,(j-1)})$ and $(\kappa^*,\xi^*)\sim q(\cdot|\kappa^{(j-1)},\xi^{(j-1)})$ that have to be chosen to achieve an algorithm that efficiently explores the posterior parameter space.  

In Sections \ref{sec:ou} and \ref{sec:ou_neuronal} we employ the generalized Adaptive Metropolis (AM) algorithm \citep{andrieu2008tutorial} to tune the two proposal distributions. Regarding the generation of proposals $\phi^{i*}$, in the first step of the blocked Gibbs scheme we tune subject-specific proposal distributions, separately for each $\phi^{i*}$. In addition to these $M$ proposal distributions we also tune a proposal distribution for $(\kappa^*,\xi^*)$. Thus, we automatically tune overall $M+1$ proposal distributions via the generalized AM algorithm. Additionally, in Sections~\ref{sec:ou} and \ref{sec:ou_neuronal} we found that the use of different proposal distributions for each $\phi^{i*}$ was 
beneficial since random effects for the different subjects varied around very different values. 



\section{Applications} \label{sec:app}

\subsection{Ornstein-Uhlenbeck SDEMEM} \label{sec:ou}

We consider the following Ornstein-Uhlenbeck (OU) SDEMEM 

\begin{align} \label{eq:ou_model}
    \left\{
        \begin{array}{ll}
        Y^i_t &= X_t^i + \epsilon_t^i, \quad \epsilon^i_t \indep \textrm{N}(0, \sigma_{\epsilon}^2), \qquad i=1,...,M \\
        d X^i_{t} &= \theta_1^i  (\theta^i_ 2 - X_t^i) dt + \theta_3^i dW_t^i.
        \end{array}
    \right.
\end{align}
Here $\theta_2^i\in\mathbb{R}$ is the stationary mean for the $\{X_t^i\}$ process, $\theta_1^i>0$ is a growth rate (expressing how rapidly the system reacts to perturbations) and $\theta_3^i$ is the diffusion coefficient.
The OU process is a standard toy-model in that it is completely tractable, that is the associated SDE has a known (Gaussian) transition density, e.g. \cite{Fuchs_2013}. This fact, coupled with the assumption that the $Y_t^i|X_t^i$ are conditionally Gaussian and linear in the latent states, implies that we can apply the Kalman filter to evaluate the likelihood function exactly. Therefore, exact inference is possible for the OU SDEMEM (both maximum likelihood and Bayesian). 
For all units $i$ we simulate $n = 200$ observations, with constant observational time-step $\Delta t$. In our setup, all random effects $(\theta_1^i,\theta_2^i,\theta_3^i)$ are assumed strictly positive, and therefore  we work with their log-transformed version and set $\phi^i = (\log \theta^i_1, \log \theta^i_2, \log \theta^i_3)$, where
\begin{align*}
\phi^{i}_{j}|\eta \indep N(\mu_j,\tau_j^{-1}), \qquad j=1,2,3
\end{align*}
and $\eta = (\mu_1,\mu_2,\mu_3, \tau_1, \tau_2, \tau_3)$, with $\tau_j$ the precision of $\phi_j^i$. 
The SDEMEM \eqref{eq:ou_model} has no parameters $\kappa$ that are shared among subjects, and the full set of parameters that we want to infer is $(\mu_1,\mu_2,\mu_3, \tau_1, \tau_2, \tau_3, \sigma_{\epsilon})$. 

As already mentioned, we can compute the likelihood  $\pi(y|\phi, \sigma_{\epsilon}) =  \prod\limits_{i=1}^M\pi(y^i|\phi^i, \sigma_{\epsilon})$ exactly, using a Kalman filter (see \cite{tornoe2005stochastic} and \cite{donnet2013review} for a description pertaining SDEMEMs). The filter can then be used in Algorithm \ref{algGibbs}, that is we avoid using the particle filter (Algorithm \ref{BPF}) and replace it with the Kalman filter in Algorithm \ref{algGibbs}. Results from Algorithm \ref{algGibbs} when using this approach are denoted with ``Kalman''. 
The transition density for the latent state is known and therefore we do not need to use an Euler-Maruyama discretization when propagating the states forward in the particle filter. Instead we propagate the particles using the simulation scheme induced by the exact transition density:
\begin{equation}
    X^i_{t + \Delta t} =  \theta_2^i + (X^i_{t} - \theta_2^i)e^{-\theta_1^i \Delta t} + \sqrt{\frac{\theta_3^{i^2}}{2\theta^i_1}(1-e^{-2\theta^i_1\Delta t})}\times u^i_t,\label{eq:OU-exact}
\end{equation}
where $u^i_t \sim N(0, 1)$ independently for all $t$ and all $i$. Clearly, the $u_t^i$ appearing in \eqref{eq:OU-exact} are among the variates that we will correlate, when implementing CPMMH, in addition to the variates produced in the resampling steps.

We compare ``Kalman'' to four further methods: ``naive  PMMH'', where we employ Algorithm \ref{algGibbs} with the naive Gibbs scheme (see Section \ref{sec:gibbs}), ``PMMH'', which is Algorithm \ref{algGibbs}, ``CPMMH-099'', which is Algorithm \ref{algGibbs} with a Crank-Nicolson proposal for the $u^i$ using a correlation of $\rho = 0.99$, and ``CPMMH-0999'' where we use a correlation of $\rho = 0.999$. The number of particle used for each method was selected using the methods described in Section \ref{sec:tuning}. All five methods return exact Bayesian inference, and while this is obvious for ``Kalman'', we remind the reader that this holds also for the other four approaches as these are instances of the pseudo-marginal approach. Therefore, special interest is in efficiency comparisons between the last four algorithms, ``Kalman'' being the obvious gold-standard.

\begin{figure}[h]
\centering
\includegraphics[scale = 0.5]{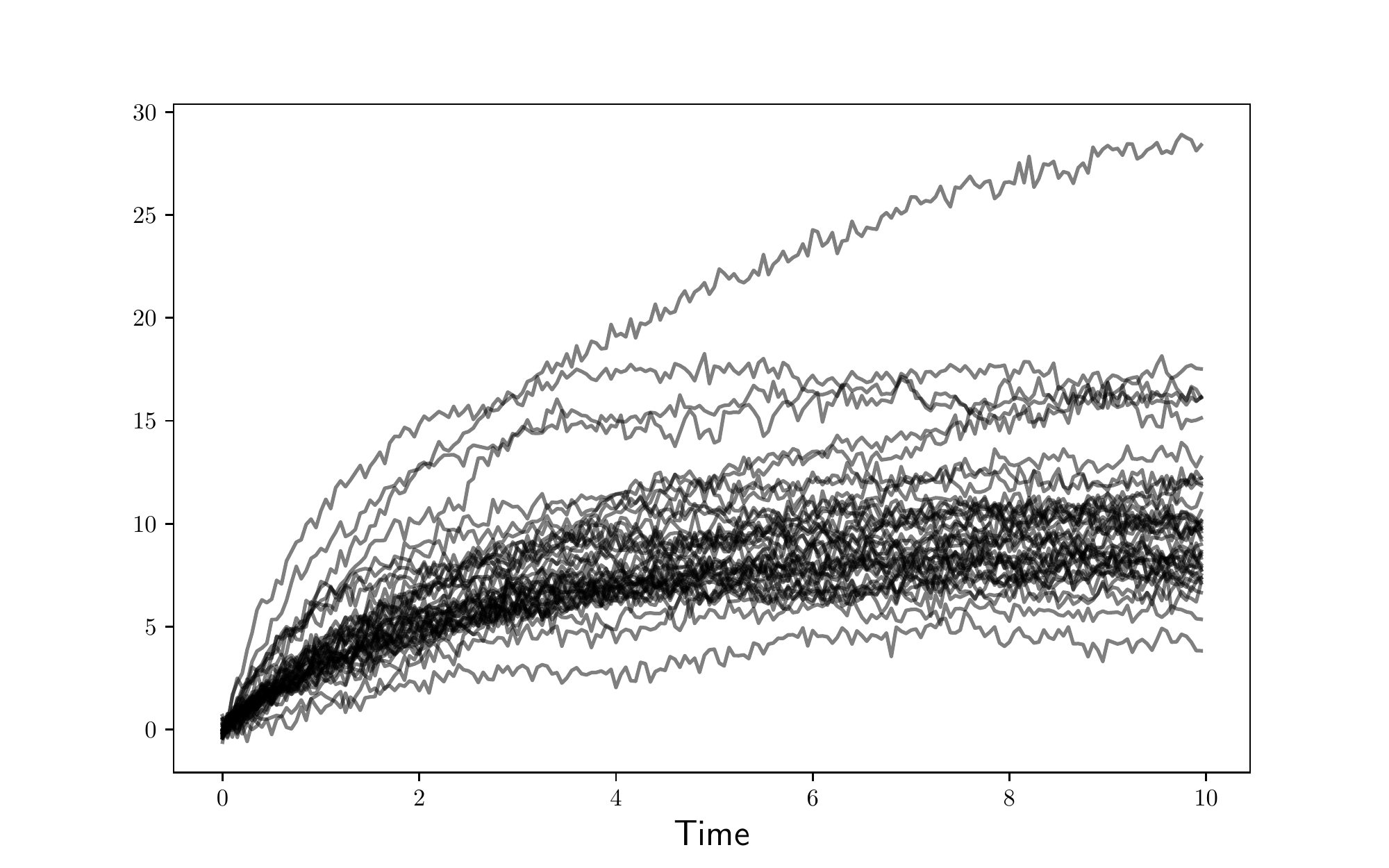}
\caption{Simulated data from the OU-SDEMEM model.}
\label{fig:ou_sim_data}
\end{figure}

We simulated data from the model in \eqref{eq:ou_model} with the following settings (data are in Figure \ref{fig:ou_sim_data}): $M=40$ experimental units,  $n = 200$ observations for each unit using a time step $\Delta t = 0.05$,  $\sigma_{\epsilon} = 0.3$,  and  $\eta = (\mu_1,\mu_2,\mu_3, \tau_1, \tau_2, \tau_3) = (-0.7, 2.3, -0.9, 4, 10, 4)$. 
The prior for the observational noise standard deviation $\sigma_{\epsilon}$ was set to a Gamma distribution $Ga(1, 0.4)$, and the priors for the $\eta$ parameters were set to

\begin{align}
    \begin{cases}
    \mu_j | \tau_j \indep \textrm{N}(\mu_{0_j}, M_{0_j} \tau_j), \qquad j=1,2,3, \label{eq:semi-conjugate_priors_ou1}\\ 
    \tau_j \indep Ga(\alpha_j, \beta_j), 
    \end{cases}
\end{align}
where, 
\begin{align*}
    (\mu_{0_1}, M_{0_1},\alpha_1, \beta_1) &= (0, 1, 2, 1), \\
    (\mu_{0_2}, M_{0_2},\alpha_2, \beta_2) &= (1, 1, 2, 0.5), \\
    (\mu_{0_3}, M_{0_3},\alpha_3, \beta_3) &= (0, 1, 2, 1). 
\end{align*}
The priors in \eqref{eq:semi-conjugate_priors_ou1} are semi-conjugate and we can therefore use a tractable Gibbs step to sample $\eta$ in step 4 of Algorithm \ref{algGibbs}. An extended introduction to the semi-conjugate prior, including the tractable posterior can be found in \cite{Murphy07conjugatebayesian}. 

\begin{figure}[h]
\centering
\includegraphics[scale=0.6]{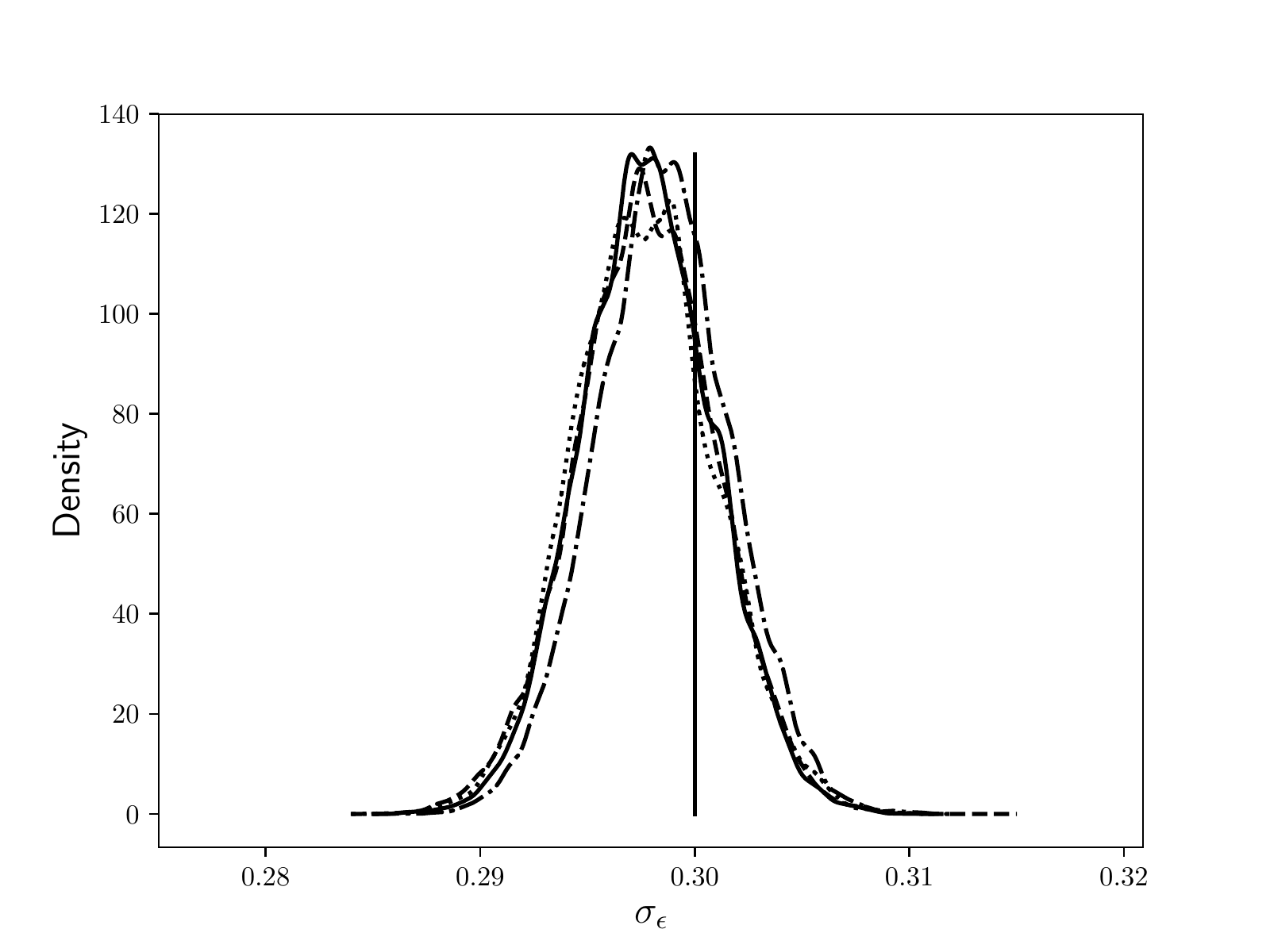}
\caption{OU SDEMEM: marginal posterior distributions for $\sigma_{\epsilon}$. Solid line is Kalman, dashed line is PMMH-naive, dotted line is PMMH, dash-dotted line is CPMMH-099, vertical line is the ground truth.}
\label{fig:ou_mp_sigma_epsilon}
\end{figure}

\begin{figure}[h]
\centering
\includegraphics[scale = 0.45]{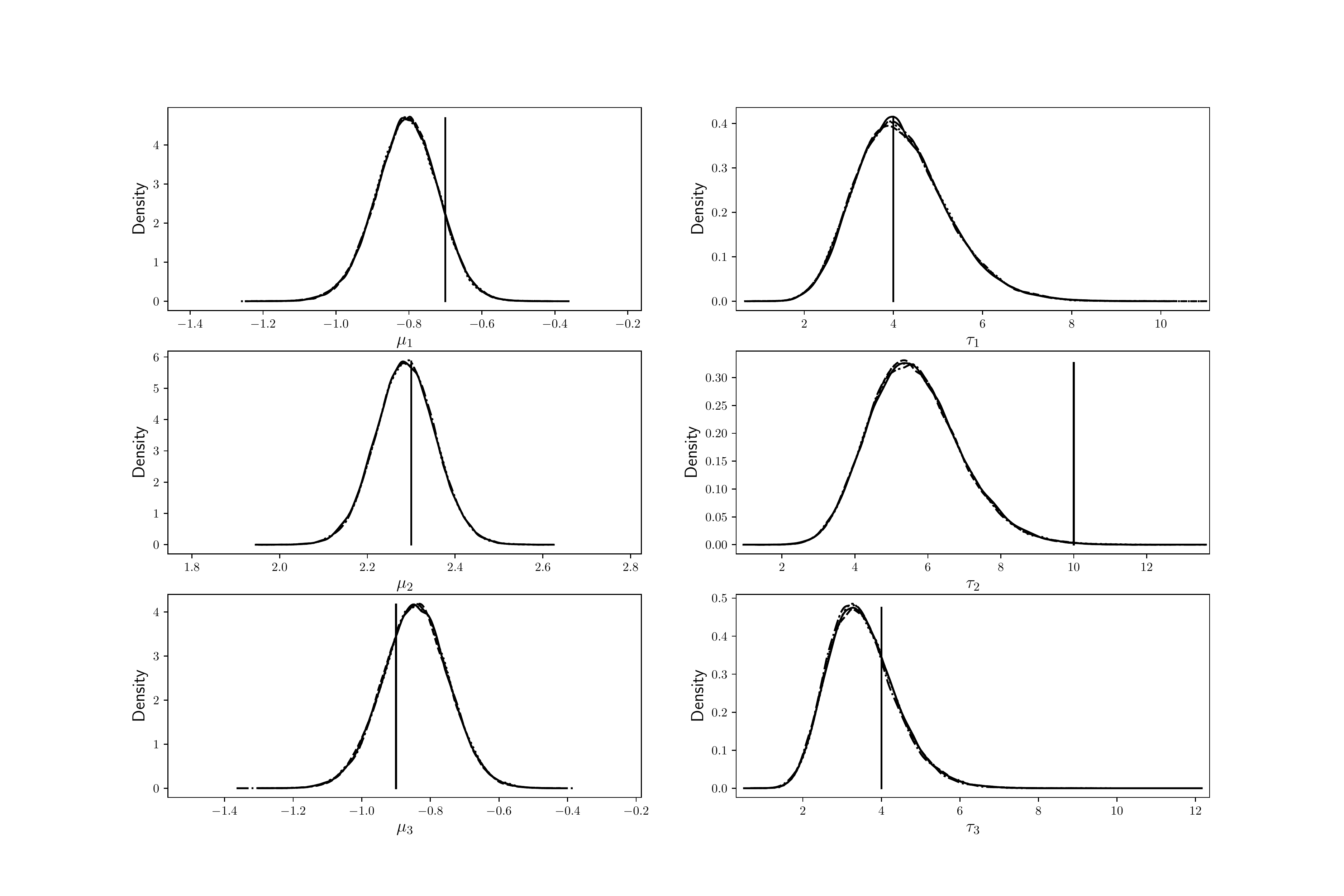}
\caption{OU SDEMEM: marginal posterior distributions for $\eta = (\mu_1,\mu_2,\mu_3, \tau_1, \tau_2, \tau_3)$. Solid line Kalman, dashed line PMMH-naive, dotted line PMMH, dash-dotted line CPMMH-099, vertical line ground truth.}
\label{fig:ou_mp_eta}
\end{figure}

We ran all four methods for $60$k iterations, considering the first $10$k iterations to be the burn-in period. We set the starting value for $\sigma_{\epsilon}$ at $\sigma_{\epsilon_0} = 0.2$, which is  far from its ground truth value. The starting values for the random effects $\phi^i_j$ were set to their prior means. The proposal distributions were adaptively tuned using the generalized AM algorithm and the particle filters were implemented on a single-core computer, thus no parallelization was utilized. We used the same number of particles $N_i\equiv N$ for all units. 
Results are in Table \ref{tab:tab_ou_sdemem} and Figures \ref{fig:ou_mp_sigma_epsilon}-\ref{fig:ou_mp_eta}. As a reference for the efficiency of the considered samplers, we take the minimum ESS per minute (mESS/m in Table \ref{tab:tab_ou_sdemem}) as measured on PMMH-naive as ``base/default'' value and set it to 1 in the rightmost column of Table \ref{tab:tab_ou_sdemem}. The minimum ESS per minute for the other samplers are relative to the PMMH-naive value. The mESS value is computed over all parameter chains (including individual random effects), i.e. the chains for $\phi$, $\sigma_\epsilon$ and $\eta$.
From Table \ref{tab:tab_ou_sdemem} we conclude that CPMMH is about 20 to 40 times more efficient than PMMH in terms of mESS/m, depending on which correlation level we use. Furthermore, ``Kalman'' is about $5140$ times more efficient than PMMH. However, the latter comparison is not very interesting since the Kalman filter can be applied only to a very restricted class of models.
The marginal posteriors in Figure \ref{fig:ou_mp_sigma_epsilon}--\ref{fig:ou_mp_eta} show that the several methods generate very similar posterior inference, which is reassuring. We left out the inference results from CPMMH-0999 for reasons of clarity. However we observed that with $N=50$ CPMMH-0999 produces a slightly biased inference for $\sigma_\epsilon$, due to failing to adequately mix over the auxiliary variable $u$, while inference for the remaining parameters is similar to the other considered methods. We verified (results not shown) that using $N=100$ is enough to repair this problem.
From Figure \ref{fig:ou_mp_sigma_epsilon}--\ref{fig:ou_mp_eta} we can conclude that all parameters, with the possible exclusion of $\tau_2$, 
are well inferred. 
Regarding $\tau_2$, this is the precision for $\theta_2^i$, the latter representing the stationary mean for a OU model. Clearly, by looking at Figure \ref{fig:ou_sim_data}, the occasional outlier in the upper part of the Figure may contribute to  underestimating the true precision of the stationary mean.
To check if CPMMH indeed is necessary, we tried to run PMMH with 100 particles (i.e., the same number of particles as for CPMMH-099). The inference results produced with PMMH with 100 particles gave considerable mismatch (in terms of posterior output) for both the $\eta$ parameters and $\sigma_{\epsilon}$ relative to that obtained from CPMMH-099, resulting from the extremely poor mixing of the chain.  

In summary, CPMMH is able to return reliable inference with a much smaller number of particles than PMMH, while resulting in a procedure that is about 20 to 40 times more efficient than PMMH (the 40-times figure is valid if we are ready to accept a small bias in $\sigma_\epsilon$). Again, for most models exact inference based on a closed-form expression for the likelihood function is unavailable, therefore being able to obtain accurate inference using a computationally cheaper version of PMMH is very appealing. 

Notice that while for this simple case study PMMH-naive has the same mESS than PMMH, this is not the case for the case study in Section \ref{sec:tum}, where using the blocked-Gibbs sampler produces a much larger mESS value compared to naive-Gibbs.
 
\begin{table}[h]
  \centering
  \small
  
  \begin{tabular}{@{}lrrrrrr@{}}
    \toprule
    Algorithm   & $\rho$ &    $N$ & CPU (m) & mESS & mESS/m & Rel. \\
    \midrule
    Kalman  & -  &- &   1.23 & 443.27 & 357.61 & 5140.18 \\
    PMMH-naive  & 0  &3000    &  4601.87 & 229.01  & 0.05   & 1.00 \\
    PMMH  & 0  &3000 &  4086.91 & 232.94  & 0.06   & 1.16 \\
    CPMMH-099  & 0.99  & 100  & 200.37 & 234.54  & 1.17  & 23.58 \\
    CPMMH-0999  & 0.999  &  50  & 110.88 & 235.63 &   2.13 & 41.48
    \\
    \bottomrule
  \end{tabular}
  \caption{OU SDEMEM. Correlation $\rho$, number of particles $N$, 
    CPU time (in minutes $m$), minimum ESS (mESS), minimum ESS per minute (mESS/m) and relative minimum ESS per minute (Rel.) as compared to PMMH-naive. All results are based on $50$k iterations of each scheme, and are medians over 5 independent runs of each algorithm on different data sets. We could only produce 5 runs due to the very high computational cost of PMMH.
    } \label{tab:tab_ou_sdemem}
\end{table}

\subsubsection{Investigating the choice of number of particles}

\begin{figure}[h]
\centering
\includegraphics[scale = 0.45]{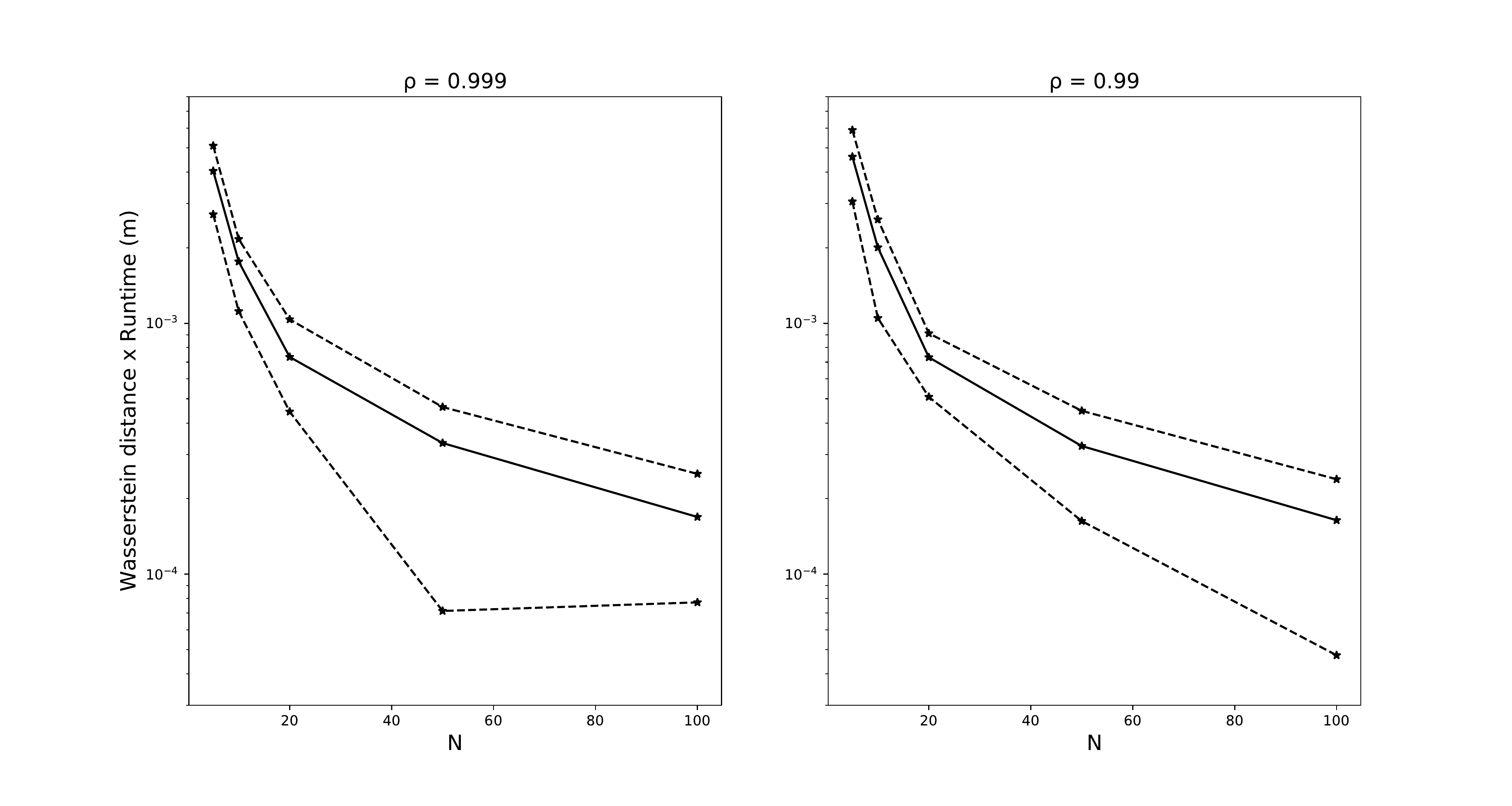}
\caption{OU SDEMEM: \textit{Wasserstein distance $\times$ runtime (m)} performance measure for the marginal posterior of $\sigma_\epsilon$, for several values of $N$ and using $\rho=0.999$ (left) and $\rho=0.99$ (right). The solid line represents the mean value obtained from the 25 different data sets. The dashed confidence bands represent the 25th and 75th percentiles. }
\label{fig:emd_runtime_sigma_eps}
\end{figure}

\begin{figure}[h]
\centering
\includegraphics[scale = 0.45]{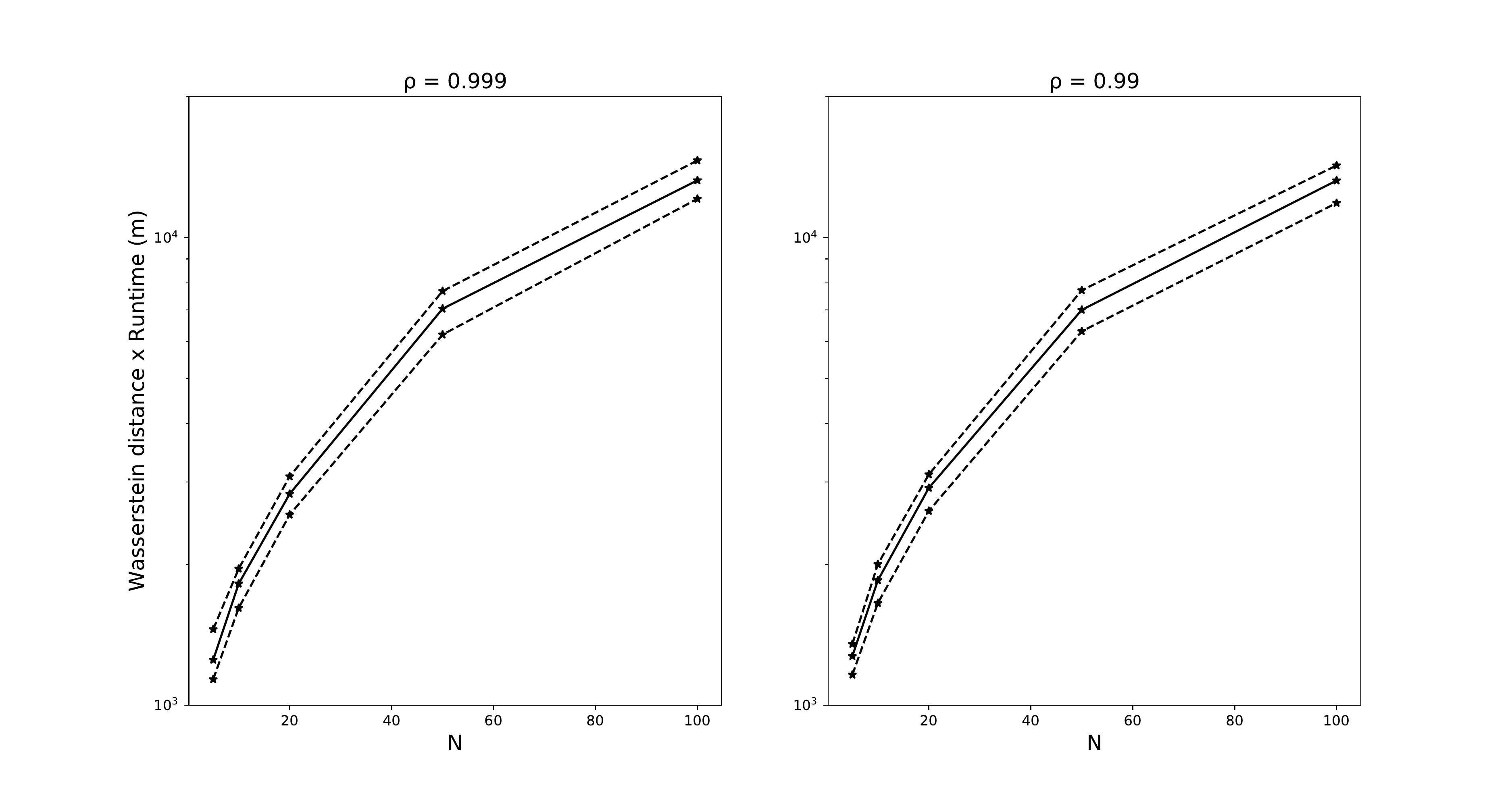}
\caption{OU SDEMEM: \textit{Wasserstein distance $\times$ runtime (m)} performance measure for the marginal posterior of $\eta$, for several values of $N$ and using $\rho=0.999$ (left) and $\rho=0.99$ (right). The solid line represents the mean value obtained from the 25 different data sets. The dashed confidence bands represent the 25th and 75th percentiles. }
\label{fig:emd_runtime_eta}
\end{figure}

A crucial problem when running methods based on particle filters is the selection of the number of particles $N$. In this section we investigate this problem by running CPMMH-099 and CPMMH-0999 with $N = [5,10,20,50,100]$ particles using 25 different (simulated) data sets. We also ran the Kalman algorithm using the 25 different data sets for comparison purposes. In this analysis, we are only interested in investigating the quality and computational efficiency of the inference. Hence, we initialised all algorithms at the ground truth parameter values and ran each algorithm for 60k iterations, and discarding the first 10k iterations as burnin period. We first estimated the Wasserstein distance, between the marginal posteriors for $\sigma_\epsilon$  and $\eta$ from the CPMMH algorithms and the corresponding Kalman-based marginal posteriors. This distance was computed via the POT package \cite{flamary2017pot} (we do not compute the Wasserstein distance for the marginal posterior of the random effects $\phi^i$, since this is not of central interest for us). All Wasserstein distances are based on the last 5k samples of the corresponding chains.  To obtain a performance measure that takes into account both the quality of the inference and the computational effort, we multiply the Wasserstein distances by the runtimes (in minutes) of the CPMMH algorithms, and obtain the performance measure \textit{Wasserstein distance $\times$ runtime (m)}; see Figure \ref{fig:emd_runtime_sigma_eps} and \ref{fig:emd_runtime_eta}. Smaller values of this measure are to be preferred as they indicate high computational efficiency and/or accurate inference. The reason for considering this performance measure is to take the quality of the inference into account, since for $N<20$ we noticed that it is possible to obtain chains that do not indicate adequate convergence  within a reasonable time-frame. 

We can conclude that, on average, results  for different correlation levels are similar. However, for $\sigma_\epsilon$ we obtain a better performance when using more particles (lower \textit{Wasserstein distance $\times$ runtime (m)}  value), this resulting from inaccurate  inference for $\sigma_\epsilon$ when using too few ($N < 50$) particles, leading to a large Wasserstein distance. However, this is not the case for $\eta $ since Figure \ref{fig:emd_runtime_eta} shows that the performance is better with fewer particles, a result that we obtain since the inference for $\eta$ is good even when using few particles (though not reported, in our analyses we observed that the Wasserstein distances for $\eta$ are similar across all attempted values of $N$). Thus, if we want to infer the measurement noise parameter $\sigma_\epsilon$ accurately, in this case we will have to use $N \geq 50$ particles, while the inference for $\eta$ is satisfactory, even with fewer particles.  

Another issue that we analyse is the variability of mESS for the different data sets, based on 50k iterations of CPMMH. To investigate this we computed the 25th and 75th percentiles of mESS for CPMMH-099 with $N=100$ and CPMMH-0999 with $N=50$ based on the inference results on all unknown parameters from 25 simulated data sets. We obtain that the 25th and 75th percentiles of mESS for CPMMH-099 ($N=100$) are $[227, 240]$, and for CPMMH-0999 ($N=50$) are $[227, 252]$. Given that the several mESS are computed on different datasets, some degree of variation in the measure is expected and we conclude that the observed mESS variability is fairly small. 

\subsection{Tumor growth SDEMEM}
\label{sec:tum}

We consider a stochastic differential mixed effects model that has been used to describe the tumor volume dynamics in mice receiving a treatment. Here we study a simplified version of the model in \cite{picchini17}, and is given by
\begin{align}
dX_{1,t}^{i}&=\left(\beta^i+(\gamma^i)^2/2\right)X_{1,t}^{i}dt+\gamma^i X_{1,t}^{i}dW_{1,t}^{i}\nonumber \\
dX_{2,t}^{i}&=\left(-\delta^i+(\psi^i)^2/2\right)X_{2,t}^{i}dt+\psi^i X_{2,t}^{i}dW_{2,t}^{i} \label{model}
\end{align}
for experimental units $i=1,\ldots,M$. Here, $W_{1,t}$ and $W_{2,t}$ are uncorrelated Brownian motion 
processes, $X_{1,t}^{i}$ and $X_{2,t}^{i}$ are respectively the volume of surviving tumor cells and volume of cells killed by a treatment for mouse $i$. 
Let $V_t^{i}=X_{1,t}^{i}+X_{2,t}^{i}$ denote the total tumor volume at time $t$ in mouse $i$. 
The observation model is given by
\begin{equation}\label{obsmodel}
Y_t^i=\log V_t^{i} + \epsilon_t^i,\qquad \epsilon_t^i\indep \textrm{N}(0,\sigma^2_e).
\end{equation}
Let $\phi^i=(\log\beta^i,\log\gamma^i,\log\delta^i,\log\psi^i)$. We complete the SDEMEM specification via the 
assumption that
\begin{equation}\label{randmodel}
\phi^{i}_{j}|\eta \indep \textrm{N}(\mu_j,\tau_j^{-1}), \qquad j=1,\ldots,4
\end{equation}
so that $\eta=(\mu_1,\ldots,\mu_4,\tau_1,\ldots,\tau_4)$.

We recognise that $X_{1,t}^{i}$ and $X_{2,t}^{i}$ are geometric Brownian motion processes and (\ref{model}) 
can be solved analytically to give
\begin{align}
X_{1,t}^{i}|X_{1,0}^i=x_{1,0}^i &\sim logN\left(\log(x_{1,0}^i)+\beta^i t\,,\,(\gamma^i)^2 t\right) \nonumber\\
X_{2,t}^{i}|X_{2,0}^i=x_{2,0}^i &\sim logN\left(\log(x_{2,0}^i)-\delta^i t\,,\,(\psi^i)^2 t\right)\label{SDEsol}
\end{align}
where $logN(\cdot,\cdot)$ denotes the log-Normal distribution. Despite the availability of a closed form solution to the underlying SDE model, the observed 
data likelihood is intractable, due to the nonlinear form of (\ref{obsmodel}) as a function of $\log(X_{1,t}^{i}+X_{2,t}^{i})$. Nevertheless, a tractable approximation can be found, by linearising $\log V_t^i$. The resulting linear noise approximation (LNA) is derived in \ref{sec:lna}, and in what follows, we compare inference under the gold standard PMMH to that obtained under the LNA. 

\begin{figure}[ht]
\centering
\includegraphics[width=10cm,height=7cm]{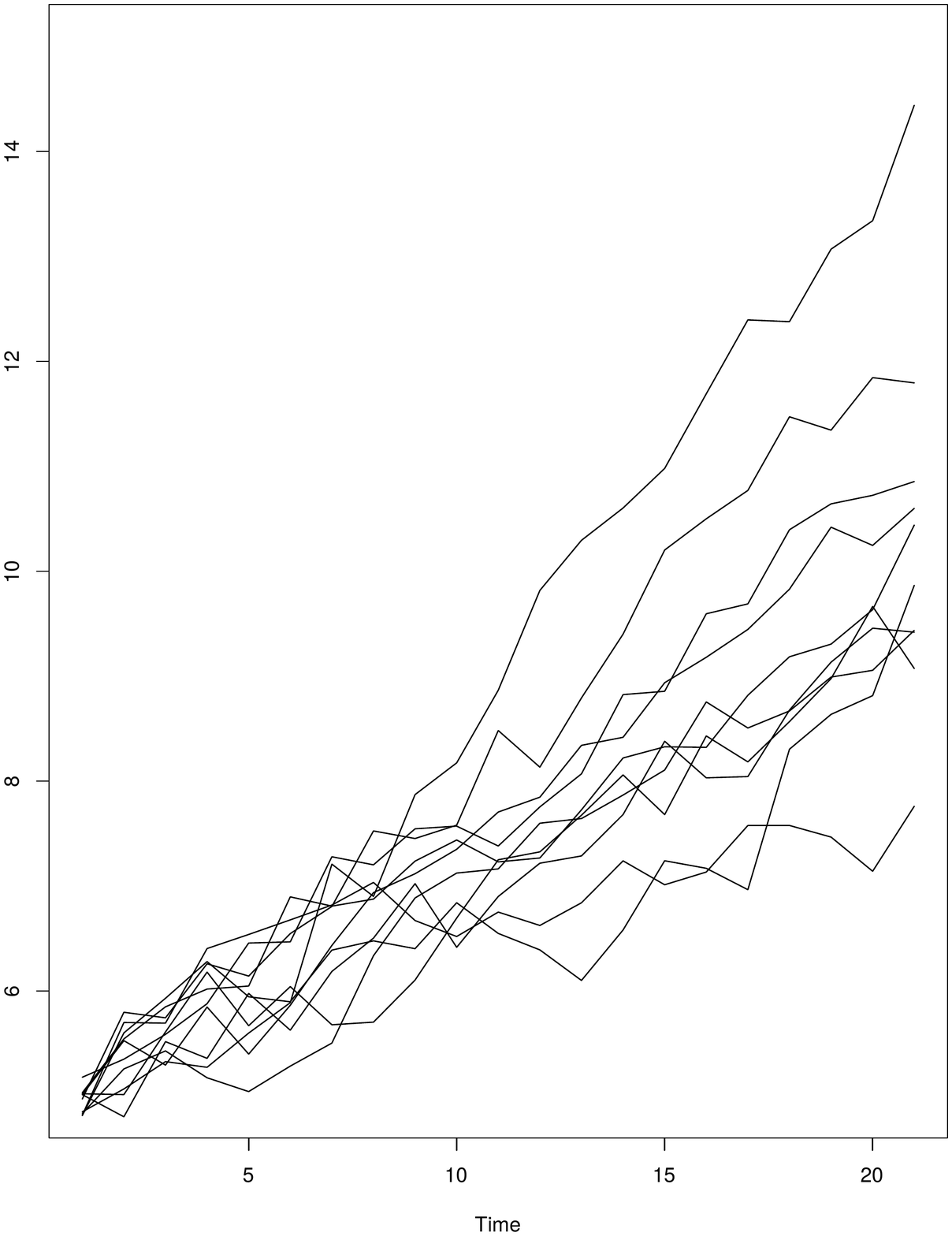}
\caption{Simulated data from the tumor growth model.}
\label{fig:simualtedTumourGrowth}
\end{figure}

\begin{figure}[ht]
\centering
\includegraphics[scale = 0.7]{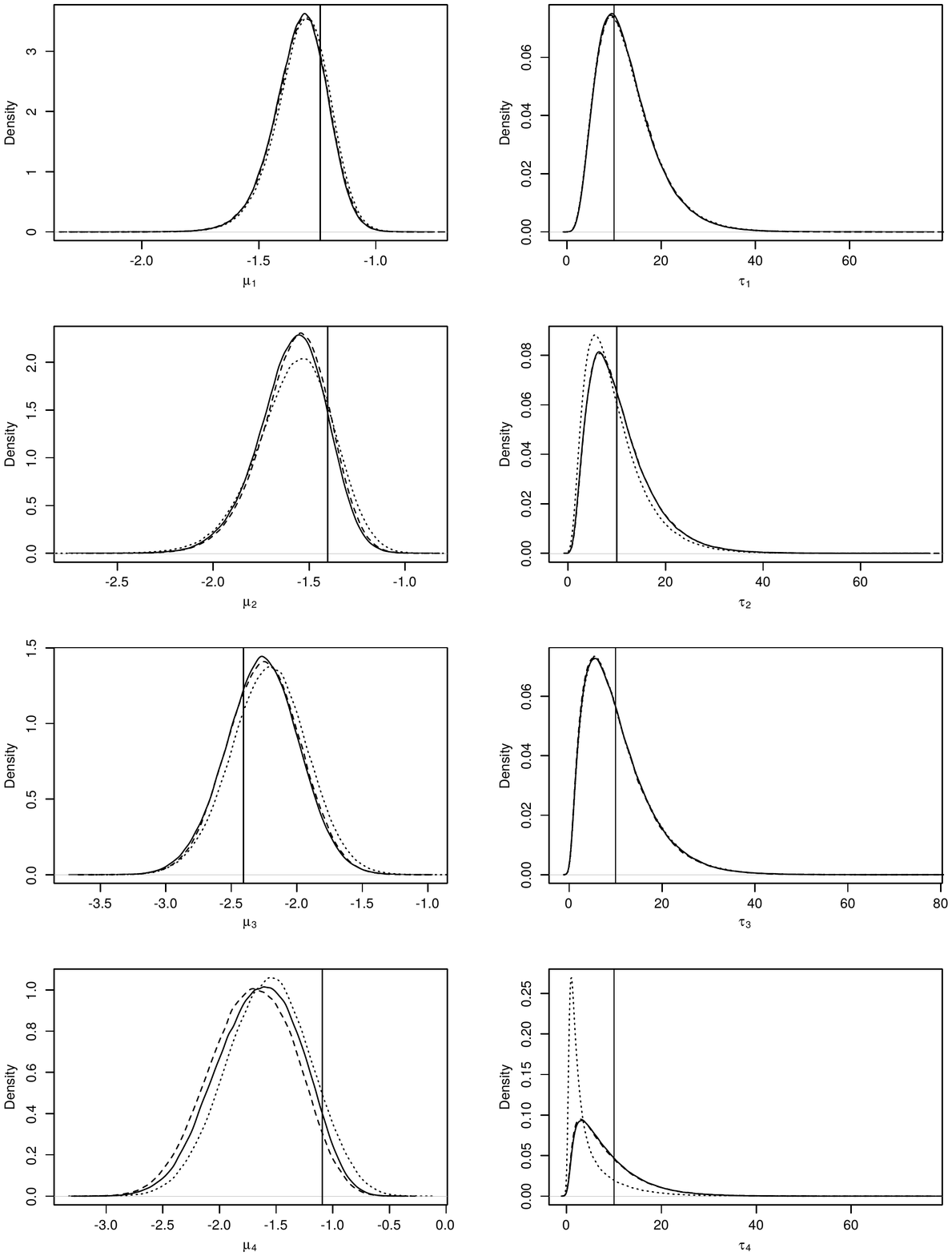}
\caption{Marginal posterior distributions for $\mu_i$ and $\tau_i$, $i=1, \ldots, 4$. Dotted line shows results from LNA scheme, solid line is from the CPMMH scheme and dashed line is the PMMH Scheme.}
\label{fig:unCorVaryUTaus}
\end{figure}

We mimicked the real data application in \cite{picchini17} by generating $21$ observations at integer times for $M=10$ replicates. We took
\[
\eta=(\log 0.29,\log 0.25,\log 0.09,\log 0.34,10,10,10,10)
\]
and sampled $\phi^{i}_{j}|\eta$ using (\ref{randmodel}). The latent SDE process was then generated using (\ref{SDEsol}) with an initial condition of $x_0^i=(75,75)^T$ (assumed known for all units), and each observation was corrupted according to (\ref{obsmodel}) with $\sigma^2_e = 0.2$. The resulting data traces are consistent with the observations on total tumor volume of those subjects receiving chemo therapy in \cite{picchini17} and can be seen in Figure ~\ref{fig:simualtedTumourGrowth}. We adopted semi conjugate, independent $\textrm{N}(-2,1)$ and $Ga(2,0.2)$ priors for the $\mu_j$ and $\tau_j$ respectively. We took $\log \sigma_e \sim \textrm{N}(0,1)$ to complete the prior specification. Given the use of synthetic data of equal length for each experimental unit, we pragmatically took the number of particles as $N_i=N$, $i=1,\ldots,10$. Our choice of $N$ was guided by the tuning advice of Section~\ref{sec:tuning}. For example, with CPMMH we obtain typical $\rho_L$ values of around 0.75, when parameter values are fixed at an estimate of the posterior mean. This gives $\sigma^2_N = 10.6$ which is achieved with $N=7$ particles. To avoid potentially sticky behaviour of the chain in the posterior tails, we choose the conservative value $N=10$. We compare four approaches: naive PMMH (where the $u^i$ are updated with both the subject specific and common parameters), PMMH (where the $u^i$ are only updated with the subject specific parameters -- Algorithm~\ref{algGibbs}), CPMMH (Algorithm~\ref{algGibbs} with a Crank-Nicolson proposal on the $u^i$) and the LNA based approach. We ran each scheme for $500k$ iterations. The results are summarised in Table~\ref{tab:tabTum} and Figure~\ref{fig:unCorVaryUTaus}.  

\begin{table}[h]
  \centering
  \small
  \begin{tabular}{@{}lrrrrrr@{}}
    \toprule
    Algorithm   & $\rho$ &    $N$ & CPU (m) & mESS & mESS/m & Rel. \\
    \midrule
    LNA           & - &-   & 1286 &3676 &2.858&13\\
    PMMH - naive  & 0  &30             &        3098  & 665 &0.215&1 \\
    PMMH  & 0  &30             & 2963  & 2559 &0.864& 4\\
    CPMMH  & 0.999  &10             &   957    & 2311 &2.415& 11\\
    \bottomrule
  \end{tabular}
  \caption{Tumor model. Correlation $\rho$, number of particles $N$, 
    CPU time (in minutes $m$), minimum ESS (mESS), minimum ESS per minute (mESS/m) and relative minimum ESS per minute (Rel.) as compared to PMMH-naive. All results are based on $500$k iterations of each scheme.}\label{tab:tabTum}
\end{table}

Figure~\ref{fig:unCorVaryUTaus} shows marginal posterior densities of the components of $\eta$. We see that inferences for these parameters are consistent with the true values that generated the data (with similar results obtained for the other parameters) and that inference via CPMMH is consistent with that from the gold-standard PMMH. Similar results are obtained for $\sigma_{\epsilon}$ (not shown for brevity). At the same time, from Table \ref{tab:tabTum} we note that CPMMH with $\rho=0.999$ is about 11 times more efficient than the naive PMMH and almost 3 times more efficient than PMMH with additional blocking. Finally, the LNA-based approach provides an accurate alternative to PMMH, except for $\tau_4$. However, everything considered, CPMMH is to be preferred here as its computational efficiency is comparable to LNA, but unlike the latter, CPMMH provides accurate inference for all parameters, and unlike LNA the CPMMH approach is plug-and-play.   

\subsubsection{Use of the Euler-Maruyama approximation}
\label{sec:tumorEM}

We anticipate that for many applications of interest, an analytic solution of the underlying SDE will not be available. It is common place to use a numerical approximation in place of an intractable analytic solution. The simplest such approximation is the Euler-Maruyama (E-M) approximation. In this section, we investigate the effect of the E-M on the performance of PMMH and CPMMH for the tumor growth model. 

The Euler-Maruyama approximation of (\ref{model}) is
\begin{align}
\Delta X_{1,t}^{i}&=\left(\beta^i+(\gamma^i)^2/2\right)X_{1,t}^{i}\Delta t+\gamma^i X_{1,t}^{i}\Delta W_{1,t}^{i}\nonumber \\
\Delta X_{2,t}^{i}&=\left(-\delta^i+(\psi^i)^2/2\right)X_{2,t}^{i}\Delta t+\psi^i X_{2,t}^{i}\Delta W_{2,t}^{i} \nonumber 
\end{align}
where, for example, $\Delta X_{1,t}^i = X_{1,t+\Delta t}^i - X_{1,t}^i$ and $\Delta W_{1,t}^i \sim N(0,\Delta t)$, with other terms defined similarly. To allow arbitrary accuracy of E-M, the inter-observation time length $\Delta t$ is replaced by a stepsize $\Delta t=1/L$ for the numerical integration, for integer $L\geq 1$. We find that using $L=5$  (giving 4 intermediate times between observation instants) allows sufficient accuracy (compared to the analytic solution)  to permit use of the same tuning choices when re-running PMMH (including the naive scheme) and CPMMH. Our findings are summarised by Table~\ref{tab:tabTumEM}.
\begin{table}[h]
  \centering
  \small
  \begin{tabular}{@{}lrrrrrr@{}}
    \toprule
    Algorithm   & $\rho$ &    $N$ & CPU (m) & mESS & mESS/m & Rel. \\
    \midrule
    PMMH - naive  & 0  &30             &        7947  & 990 &0.123&1 \\
    PMMH  & 0  &30             & 7651  & 2240 &0.293& 2.4\\
    CPMMH  & 0.999  &10             &   1893    & 2172 &1.15& 9.2\\
    \bottomrule
  \end{tabular}
  \caption{Tumor model (Euler-Maruyama). Correlation $\rho$, number of particles $N$, 
    CPU time (in minutes $m$), minimum ESS (mESS), minimum ESS per minute (mESS/m) and relative minimum ESS per minute (Rel.) as compared to PMMH-naive. All results are based on $500$k iterations of each scheme.}\label{tab:tabTumEM}
\end{table}

Unsurprisingly, inspection of Table~\ref{tab:tabTumEM} reveals that relative performance between the three computing pseudo-marginal schemes is similar to that obtained when using the analytic solution; CPMMH provides almost an order of magnitude increase in terms of mESS/m over a naive PMMH approach. We note that use of the Euler-Maruyama approximation requires computation and storage of an additional $1/\Delta t$ innovations per SDE component, inter-observation interval, particle and subject, thus accounting for the increase in CPU time compared to when using the analytic solution. Nevertheless, we find that our proposed approach is able to accommodate an intractable SDE scenario and provides a worthwhile increase in performance over competing approaches.

\subsubsection{Comparison with ODEMEM}
\label{sec:ode}

\begin{figure}[ht]
\centering
\includegraphics[width=15cm,height=15cm]{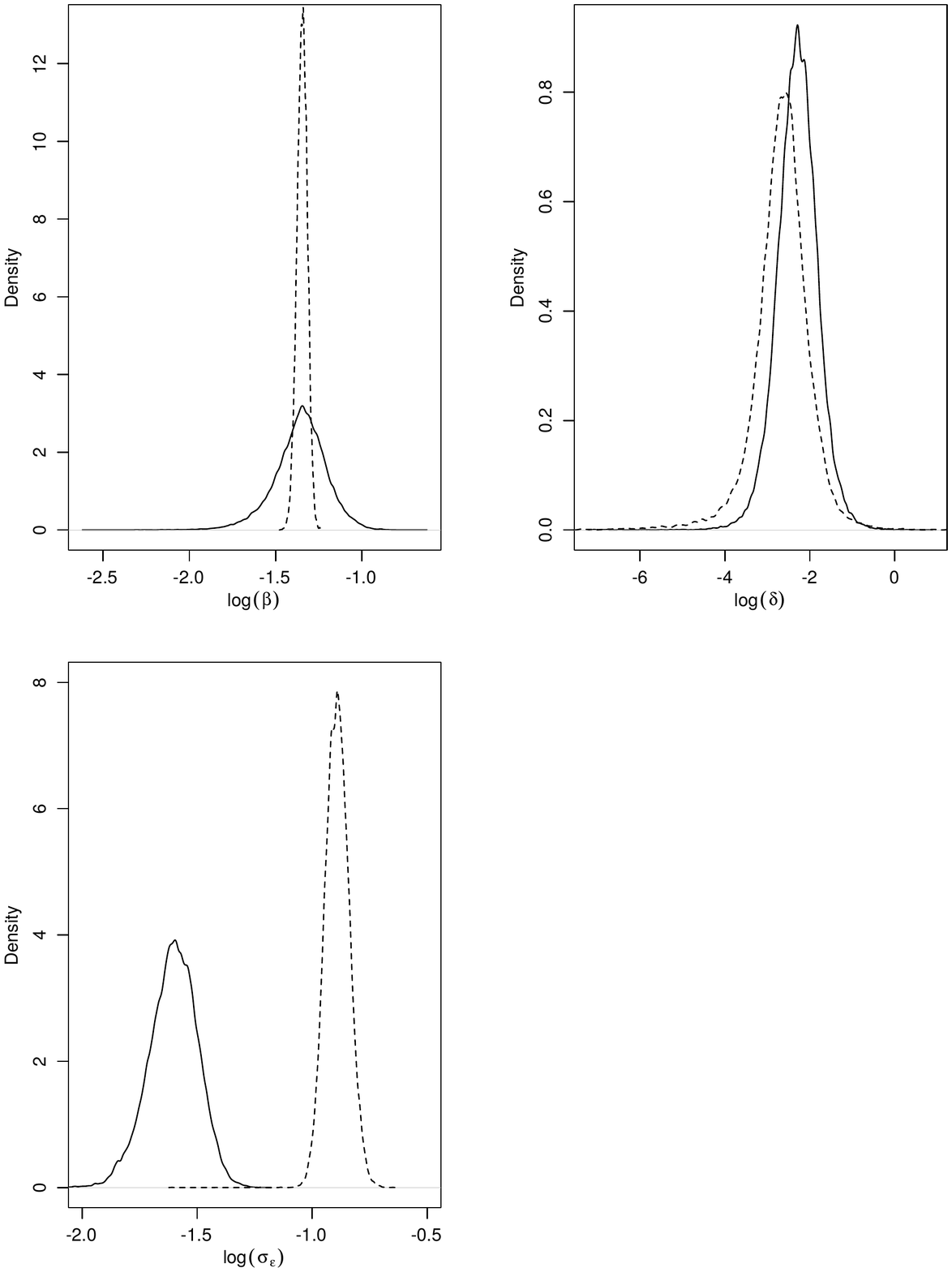}
\caption{Marginal posterior distributions for the (logged) subject specific parameters $\log\beta^1$, $\log\delta^1$, and the observation standard deviation $\log\sigma_e$. Dashed line shows results from ODEMEM, solid line is from SDEMEM.}
\label{fig:ODEvsSDEpost}
\end{figure}

\begin{figure}[ht]
\centering
\includegraphics[scale=0.55]{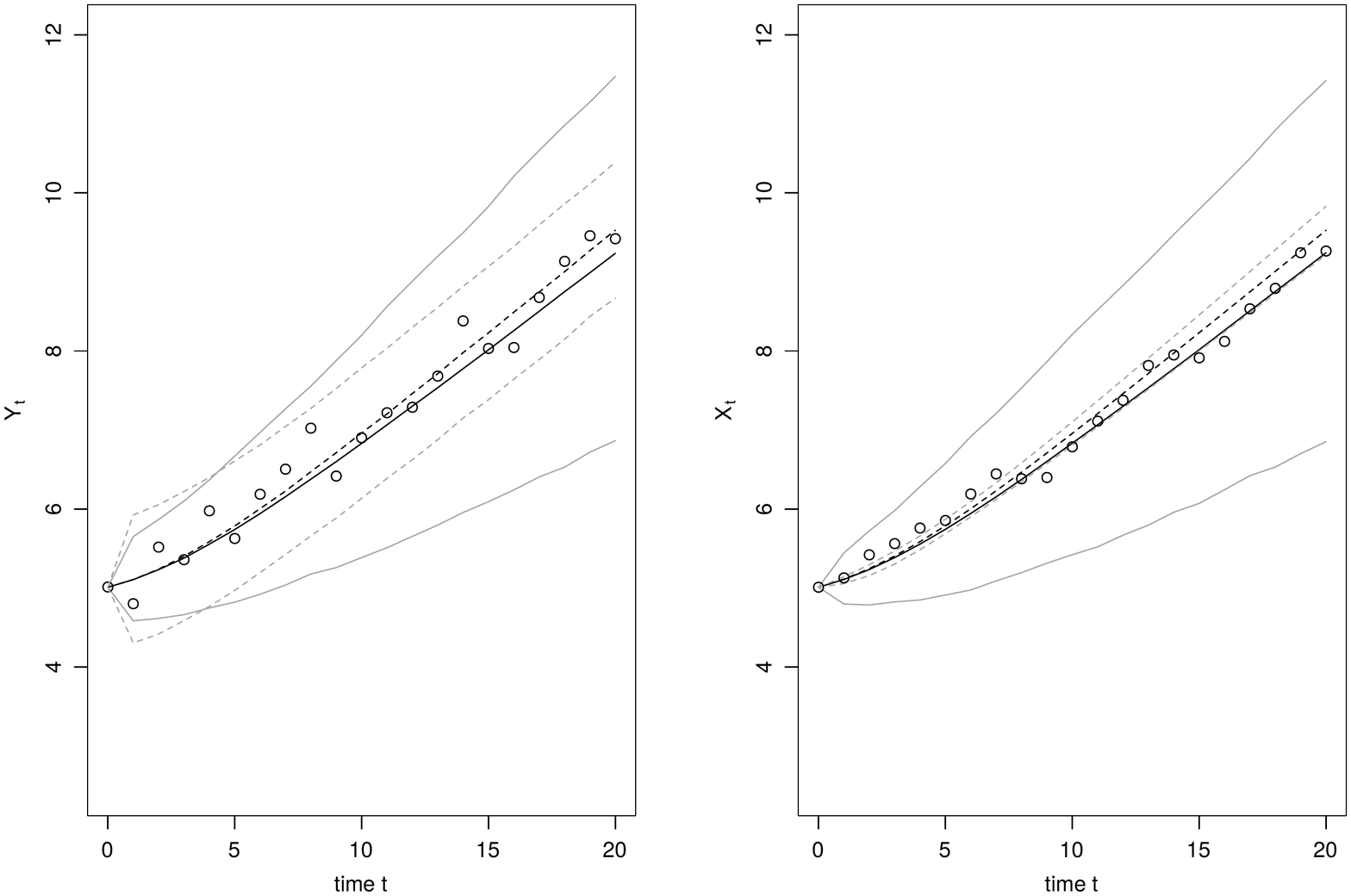}
\caption{Posterior predictive mean (black) and 95\% credible intervals (grey) for the observed process $Y_t^1$ (circles, left panel) and the latent process $X_t^1=\log V_t^1$ (circles, right panel). Dashed line shows results from ODEMEM, solid line is from SDEMEM.}
\label{fig:ODEvsSDEpostpred}
\end{figure}

To highlight the potential issues that arise by ignoring inherent stochasticity, we consider inference for an ordinary differential equation mixed effects model (ODEMEM) of tumor growth. We take the SDEMEM in (\ref{model}) and set $\gamma^i=\psi^i=0$ to give
\begin{align}
dx_{1,t}^{i}&=\beta^i x_{1,t}^{i}dt, \nonumber \\
dx_{2,t}^{i}&=-\delta^i x_{2,t}^{i}dt\label{ODEmodel}
\end{align}
for $i=1,\ldots,M$. The observation model and random effects distributions remain unchanged from (\ref{obsmodel}) and (\ref{randmodel}) upon omitting $\log\gamma^i$ and $\log\psi^i$ from $\phi^i$. 
The ODE system in (\ref{ODEmodel}) can be solved to give
\[
x_{1,t}^i=x_{1,0}^i\exp\{\beta^i t\}, \qquad x_{1,t}^i=x_{1,0}^i\exp\{\delta^i t\}.
\]
The likelihood associated with each experimental unit is then obtained simply as
\[
\pi(y^i|\phi^i,\sigma_e)=\prod_{t=1}^{21}\textrm{N}\left(y_t^i; \log(x_{1,t}^i+x_{2,t}^i),\sigma_e^2\right).
\]
Fitting the ODEMEM to the synthetic data set from Section~\ref{sec:tum} is straightforward, via a Metropolis-within-Gibbs scheme. Figures~\ref{fig:ODEvsSDEpost} and \ref{fig:ODEvsSDEpostpred} summarise our findings. Unsurprisingly, since the ODEMEM is unable to account for intrinsic stochasticity, the observation standard deviation is massively over-estimated. Figure~\ref{fig:ODEvsSDEpost} shows little agreement between the marginal posteriors under the ODEMEM and SDEMEM for this parameter. In terms of model fit, both the observation ($Y_t^1$) and latent process ($X_t^1=\log V_t^1$) predictive distributions for unit 1 are over concentrated for the ODEMEM. Similar results (not shown) are obtained for the other experimental units. Notably, from Figure~\ref{fig:ODEvsSDEpostpred}, around half of the actual simulated $X_t$ values lie outside of the 95\% credible interval under the ODEMEM.  

\subsection{Neuronal data}\label{sec:ou_neuronal}

\begin{figure}
    \centering
    \includegraphics[scale = 0.7]{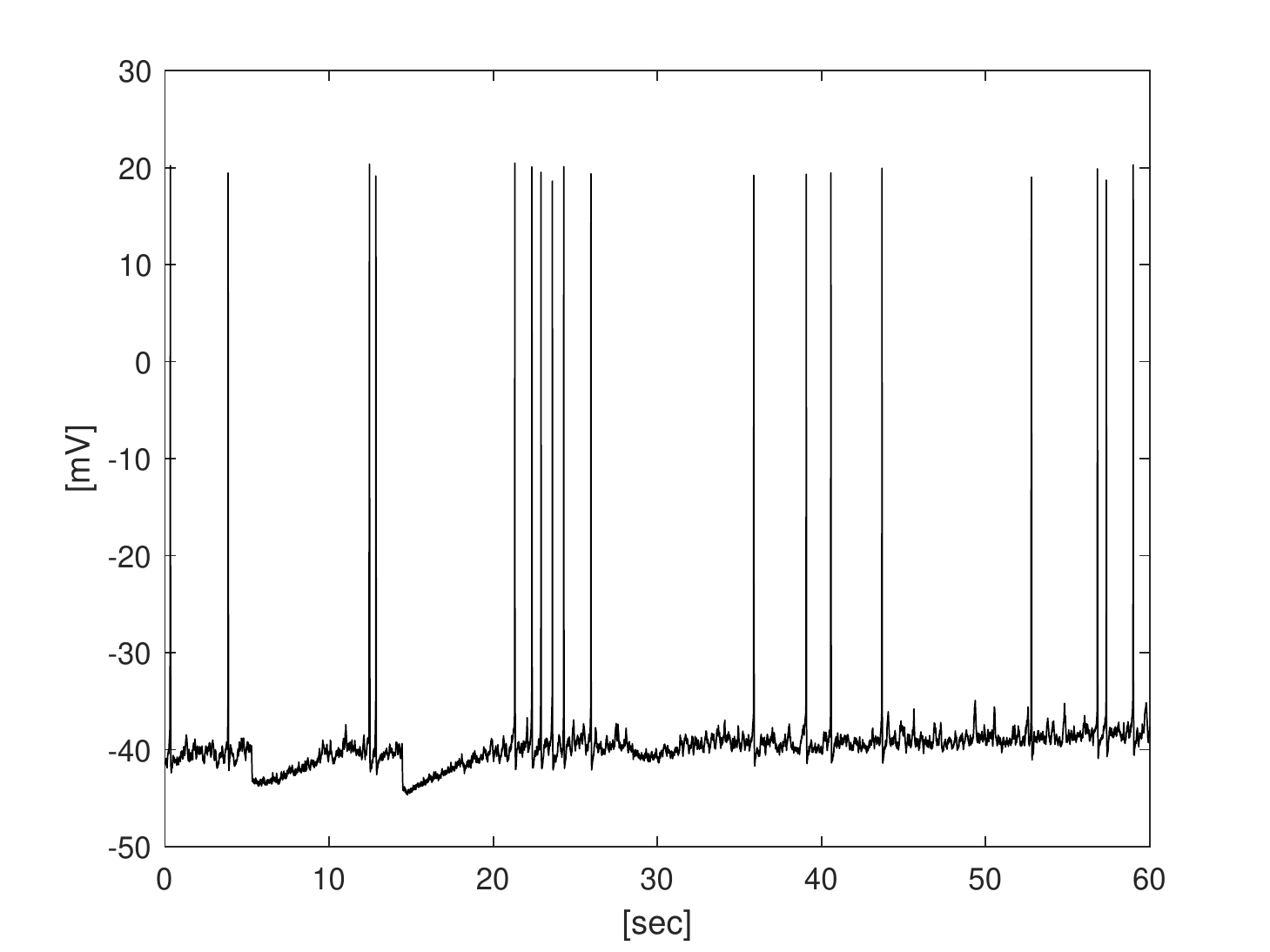}
    \caption{An exemplificative plot of depolarization [mV] vs time [sec] (data from \cite{hopfner2007set}).}
    \label{fig:hopfner}
\end{figure}

\begin{figure}[h]
\centering
\includegraphics[scale = 0.5]{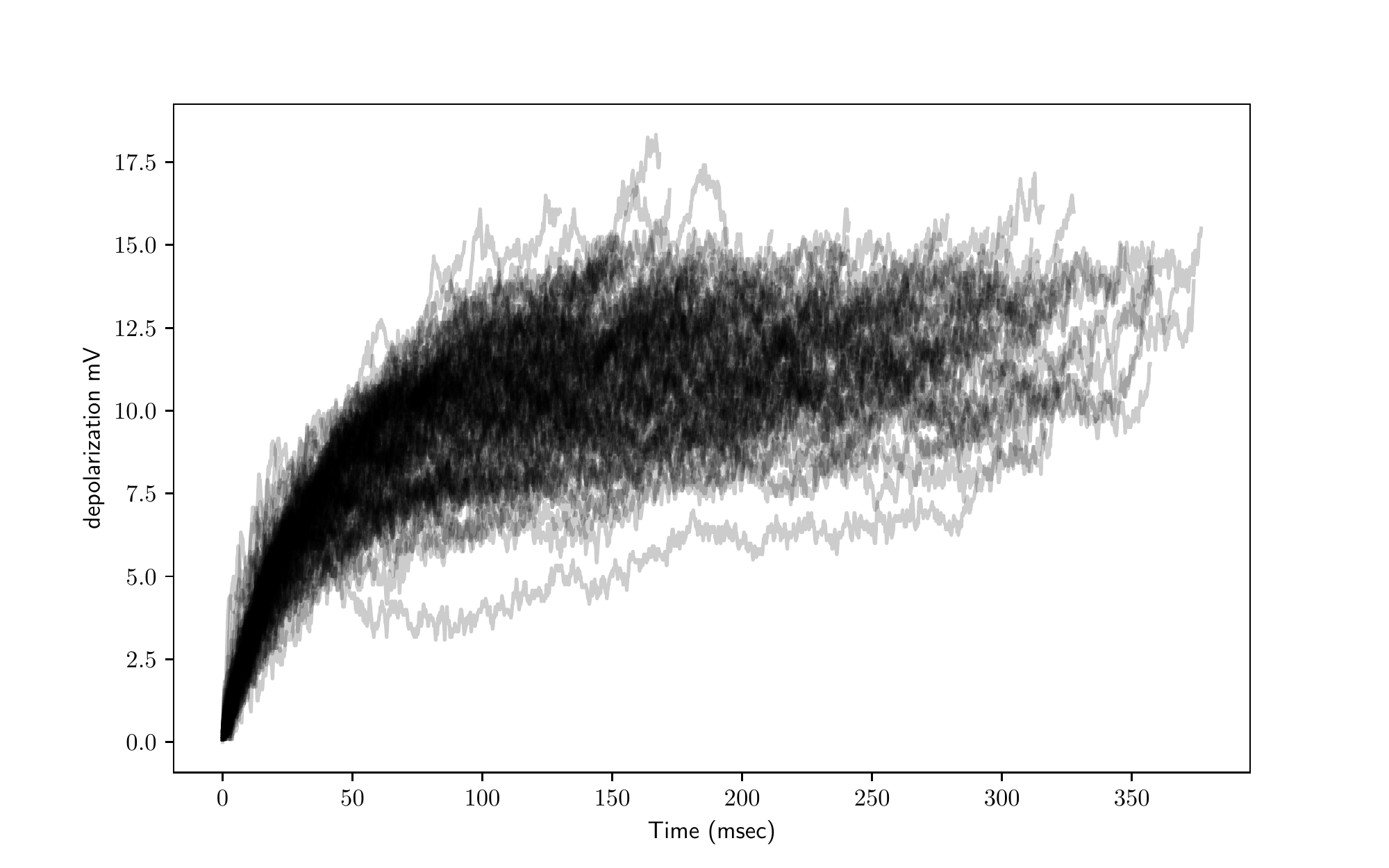}
\caption{Observations from 100 ISIs.}
\label{fig:ou_neuron_data}
\end{figure}

Here we consider a much more challenging problem: modelling a large number of observations pertaining neuronal data. In particular, we are interested in modelling the neuronal membrane potential across inter-spike intervals (ISIs). The problem of modelling the membrane potential from ISIs measurements using SDEs has already been considered numerous times, also using SDEMEMs, see \cite{picchini2008parameters}. In fact here we analyze the same data considered in \cite{lansky2006parameters} and \cite{picchini2008parameters}, or actually a subset thereof, due to computational constraints. The ``leaky integrate-and-fire'' appears to
be one of the most common models, in both artificial neural network
applications and descriptions of biological systems. Deterministic and stochastic implementations of the model are possible. In the stochastic version, under specific assumptions \citep{lansky1984approximations}, it coincides with the Ornstein-Uhlenbeck stochastic process
and has been extensively investigated in the neuronal
context, for instance in \cite{ditlevsen2005estimation}. Consider Figure \ref{fig:hopfner} as an illustrative example, reporting values of neuronal membrane depolarization studied in \cite{hopfner2007set}. Inter-spike-intervals are the observations considered between ``firing'' times of the neuron, the latter being represented by the spikes appearing in Figure \ref{fig:hopfner} (notice these are not the data we analysed. This figure is only used for illustration). Data corresponding to the near-deterministic spikes are removed, and what is left constitutes data from several ISIs. As in \cite{picchini2008parameters}, we consider data  from different ISIs as independent. Hence, $M$ is the number of considered ISIs. These are 312 in total, however, because of computational limitations, we will only analyze a subset of 100 ISIs, hence our results are based on $M=100$ and a total of 162,610 observations. A challenge is posed by the fact that some ISIs are much longer than others (in our case they vary between 600 and 2,500 observations), meaning that longer ISIs could typically require a larger $N$ to avoid particle depletion, but using the same large $N$ to approximate all $M$ likelihood terms would be a waste of computational resources. This is why CPMMH comes particularly useful, as it allows to keep a small $N$ across all units while still avoiding sticky behaviour in the MCMC chains.   
Data from the 100 ISIs are plotted on a common time-scale in Figure \ref{fig:ou_neuron_data} (after some translation to let each ISI start approximately at zero value at time zero).
These consist of membrane potentials measured every 0.15 msec intracellularly from the auditory
system of a guinea pig (for details on data acquisition and processing, see \cite{yu2004corticofugal}).  

Outside the mixed-effects context, if we denote the neuronal input with $\nu$, and if the neuron is supposed to operate in a stationary state during some time of interest, then $\nu$ would be assumed constant during this period. \cite{picchini2008parameters} generalize by assuming that in addition to $\nu$ there is a random component changing from one ISI to the next, which could be caused by the naturally occurring variations  of environment signaling, by experimental irregularities or by other sources of noise not included in the model. This fact can then be modeled by assuming that each ISI has its own input $\nu^i$, and  \cite{picchini2008parameters} specifically assume that the $\nu^i$ are iid Gaussian distributed with mean $\nu$.
An extension of the model in \cite{picchini2008parameters} is the following state-space type SDEMEM 

\begin{align} \label{eq:neurnal_data}
    \left\{
        \begin{array}{ll}
        Y^i_t &= X_t^i + \epsilon_t^i, \quad \epsilon^i_t \indep \textrm{N}(0, \sigma_{\epsilon}^2), \qquad i=1,...,M, \\
        d X^i_{t} &= (-\lambda^i X_t^i + \nu^i) dt + \sigma^i dW_t^i.
        \end{array}
    \right.
\end{align}
where the diffusion process $\{X_t^i; t \ge 0\}$ models the membrane potential [mV] in the $i$th ISI, with input $\nu^i$ $[\text{mV/msec}]$. The spontaneous voltage decay (in the absence of input) for the $i$th ISI is  $(\lambda^i)^{-1}$ $[\text{msec}]$, which means that the stationary mean for $\{X_t^i\}$ is $\nu^i/\lambda^i$, see e.g. \cite{ditlevsen2005estimation} for details. The diffusion coefficients $\sigma^i$ have unit [mV/$\sqrt{\mathrm{msec}}$]. Clearly, we assume that we are unable to observe $\{X_t^i\}$ directly, and instead can only observe a noisy realization from $\{Y_t; t \ge 0\}$. Differences with the SDEMEM in \cite{picchini2008parameters} are that: (i) their observations were assumed unaffected by measurement noise, i.e. observations were directly available from $\{X_t^i; t \ge 0\}$, $i=1,...,M$, which is a convenient assumption easing calculations towards obtaining exact maximum likelihood estimation, but that it is generally possible to argue against; (ii) in \cite{picchini2008parameters} the only random effect was $\nu^i$, and remaining parameters were fixed-effects, while in the present case we have random effects $\lambda^i$ and $\sigma^i$ in addition to $\nu^i$. Of course here we also need to estimate $\sigma_\epsilon$, which was not done in \cite{picchini2008parameters} since no measurement error was assumed.

As in Section \ref{sec:ou} the random effects are constrained to be positive and we therefore define $\phi^i = (\phi^i_1,\phi^i_2,\phi^i_3)=(\log \lambda^i, \log \nu^i, \log \sigma^i)$, where
\begin{align*}
\phi^{i}_{j}|\eta \indep \textrm{N}(\mu_j,\tau_j^{-1}), \qquad j=1,2,3,
\end{align*}
and $\eta = (\mu_1,\mu_2,\mu_3, \tau_1, \tau_2, \tau_3)$, with $\tau_j$ the precision of $\phi_j^i$.
Since we here have a similar setting as in Section \ref{sec:ou}, we employ the same semi-conjugate priors with hyperparameters
\begin{align*}
    (\mu_{0_1}, M_{0_1},\alpha_1, \beta_1) &= (\log(0.1), 1, 2, 1), \\
    (\mu_{0_2}, M_{0_2},\alpha_2, \beta_2) &= (\log(1.5), 1, 2, 1), \\
    (\mu_{0_3}, M_{0_3},\alpha_3, \beta_3) &= (\log(0.5, 1, 2, 1). 
\end{align*}
The considered data are measured with techniques ensuring high precision, and we assume the following prior $\log \sigma_{\epsilon}\sim \textrm{N}(-1, 1)$. Because of the small measurement noise, we expect that a bootstrap filter will perform poorly, leading to a very noisy approximation of the likelihood $\pi(y|\phi, \sigma_{\epsilon}) =  \prod\limits_{i=1}^M\pi(y^i|\phi^i, \sigma_{\epsilon})$. To be able to obtain a good approximation of the likelihood, we instead use the bridge particle filter found in \cite{golightly2011bayesian}, 
since, as explained below, the bootstrap filter is statistically inadequate for this experiment (moreover, it is also computationally inadequate, since it would require a too large number of particles, which was impossible to handle with the limited memory of our computer).
In \ref{sec:bridge_filter}, we derive the bridge filter for the model in \eqref{eq:neurnal_data}, and we also compare the  forward propagation of the particles that we obtain using the bootstrap filter and the bridge filter. In \ref{appendix:comp_bootstrap} we see that the likelihood approximation obtained from the bootstrap filter is very inaccurate, which is due to its inability to handle measurements with small observational noise. Consequently, the number of particles required to give likelihood estimates with low variance is computationally prohibitive. Therefore, for this example, we only report results based on the bridge filter (which is not a plug-and-play method).

\begin{figure}[ht]
\centering
\includegraphics[scale=0.6]{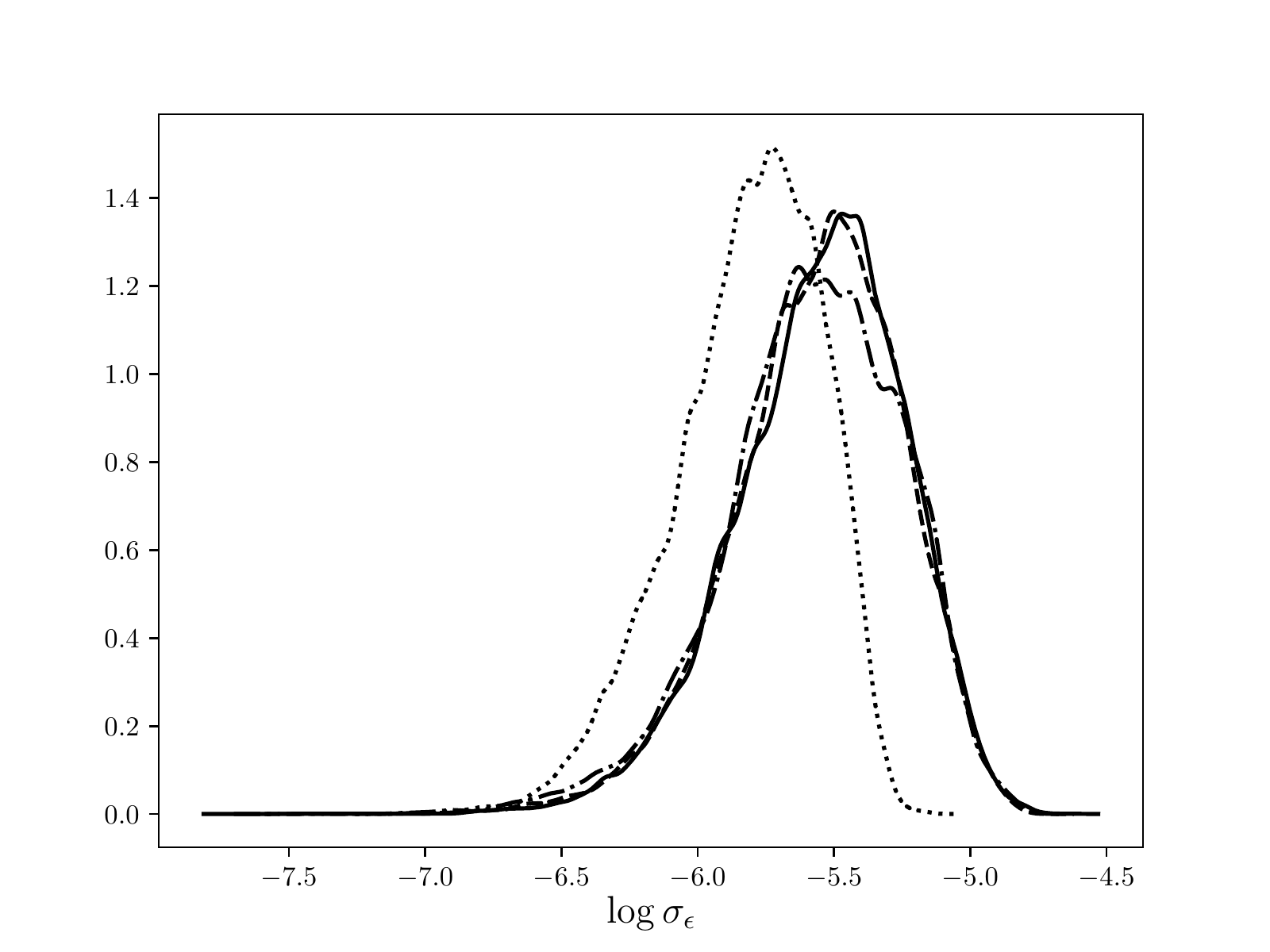}
\caption{Neuronal model: marginal posterior distributions for $\log\sigma_{\epsilon}$. Solid line is Kalman, dashed line is PMMH, dotted line is CPMMH-0999, dash-dotted line CPMMH-09.}
\label{fig:neuronal_mp_sigma_epsilon}
\end{figure}
\begin{figure}[ht]
\centering
\includegraphics[scale=0.45]{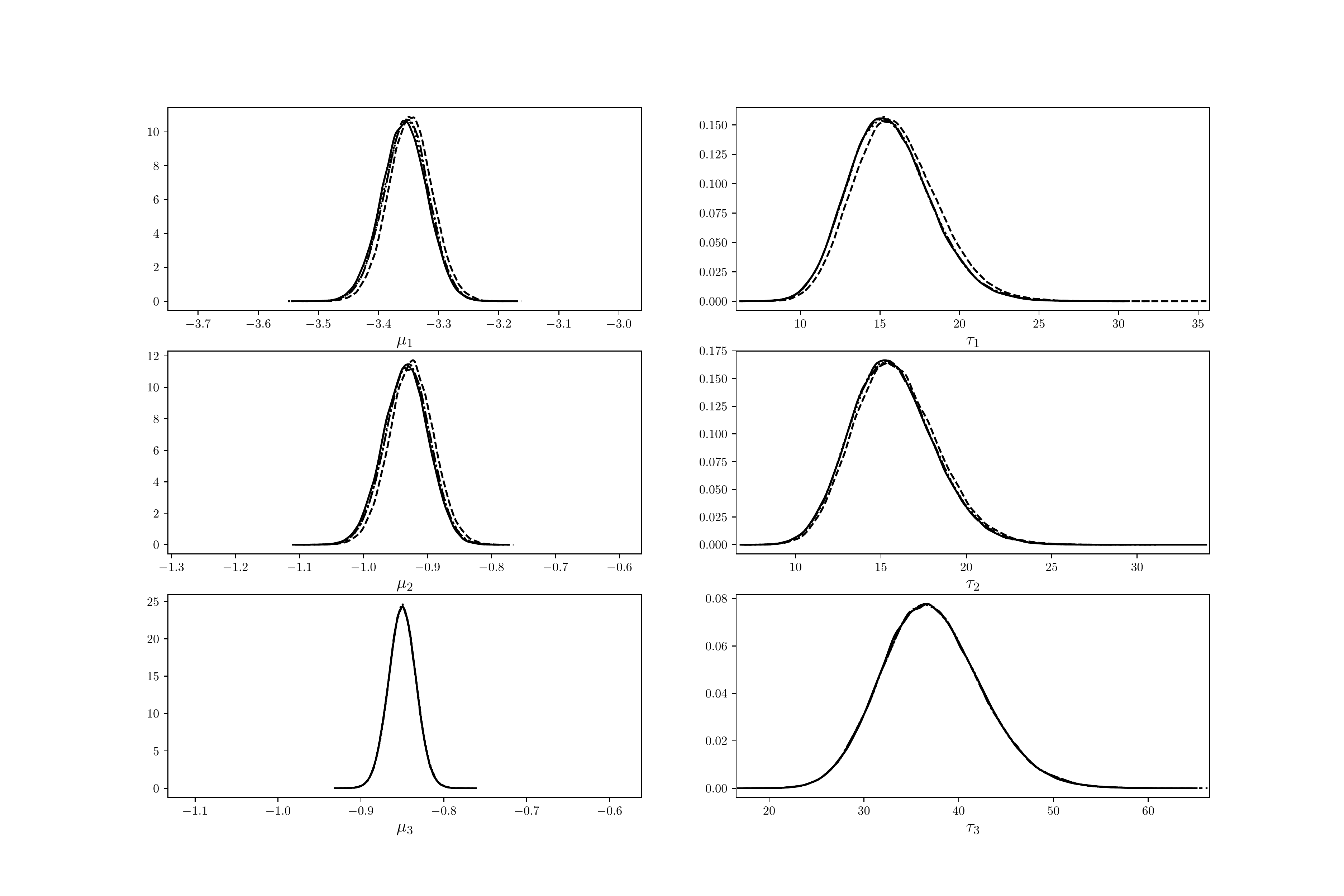}
\caption{Neuronal model: marginal posterior distributions for  $\eta = (\mu_1,\mu_2,\mu_3, \tau_1, \tau_2, \tau_3)$. Solid line is Kalman, dashed line is PMMH, dotted line is CPMMH-0999, dash-dotted line CPMMH-09.}
\label{fig:neuronal_mp_eta}
\end{figure}

We use the following four algorithms already defined in Section \ref{sec:ou}: Kalman, which obviously here is the gold-standard method; PMMH, using the bridge filter with $N=1$ particle;  CPMMH-0999  using the bridge filter also with 1 particle, and  CPMMH-09  using the bridge filter with 1 particle. 
We find that, due to propagating particles conditional on the next observation, using a single particle was enough to give likelihood estimates with low variance.
We ran all algorithms for 100k iterations, considering the first 20k iterations as burn-in. The starting value for $\sigma_{\epsilon}$ was set far away from the posterior mean that we obtained from a pilot run of the Kalman algorithm, and the starting values for the random effects $\phi^i_j$ were set to their prior means. For all algorithms, the proposal distributions were tuned adaptively using the generalized AM algorithm as described in Section \ref{sec:tune_prop_dist}. We ran the algorithms on a single-core computer so no parallelization was utilized. 
Posterior marginals in Figures \ref{fig:neuronal_mp_sigma_epsilon}-\ref{fig:neuronal_mp_eta} show that inference results for all algorithms are very similar, except for CPMMH-0999, for which posterior samples of $\sigma_{\epsilon}$ are inconsistent with the output from the other competing schemes. We note that the case of $N=1$ can be seen to correspond to a joint update of the parameters and latent process $x$. Inducing strong positive correlation between successive values of $u$ therefore results in extremely slow mixing over the latent process and in turn, the parameters. This is particularly evident for $\sigma_\epsilon$, whose update requires calculation of likelihood estimates over all experimental units.
Reducing $\rho$ to 0.9 appears to alleviate this problem.
Runtimes and ESS values are in Table \ref{tab:tab_neuronal_sdemem}. As expected, Kalman is the most efficient algorithm, being 19 times more efficient than PMMH is terms of ESS/min.  However, here PMMH and CPMMH have the same efficiency in terms of ESS/min. Thus, CPMMH does not seem to produce any efficiency improvement for this case study. This is due to the efficiency of the bridge filter in guiding state proposals towards the next observation, and therefore allowing us to run PMMH with very few particles, thus making the potential improvement brought by CPMMH essentially null.   

We compare our results with those in \cite{picchini2008parameters}. Since we have assumed that the random effects $\phi^i = (\phi^i_1,\phi^i_2,\phi^i_3)=(\log \lambda^i, \log \nu^i, \log \sigma^i)$ are Gaussian, then the $(\lambda^i, \nu^i,\sigma^i)$ are log-Normal distributed with means $(\lambda,\nu,\sigma)$ and standard deviations $(\sigma_\lambda,\sigma_\nu,\sigma_\sigma)$ respectively. By plugging the posterior means for $(\log \lambda^i, \log \nu^i, \log \sigma^i)$ as returned by ``Kalman'' into the formulas for the mean and standard deviation of a lognormal distribution, we obtain that $\lambda=0.036$ $(\sigma_\lambda=0.009)$ [1/msec], $\nu=0.406$ $(\sigma_\nu=0.105)$ [mV/msec], and $\sigma= 0.433$, $(\sigma_\sigma=0.072)$. In \cite{picchini2008parameters} we used a maximum likelihood approach, which is a fast enough procedure for Markovian data (there we did not assume a state-space model) that allowed us to obtain point estimates using all 312 ISIs (instead of 100 ISIs as in this case), but still slow enough to not permit bootstrapped confidence intervals to be obtained. Therefore, there we reported intervals based on asymptotic normality. There we had point estimates $\hat{\nu}=0.494$ and $\hat{\sigma}_\nu=0.072$, which are similar to our Bayesian estimation. It makes sense that the inferences are not very different, as in the end our estimation of $\sigma_\epsilon$ is very small, meaning that we could assume nearly Markovian data. However here we have also inferences for random effects $\lambda^i$ and $\sigma^i$, whereas in \cite{picchini2008parameters} these were assumed fixed (unknown) effects with maximum likelihood estimates $\hat{\lambda}=0.047$ [1/msec] (it can be obtained from Table 1 in \cite{picchini2008parameters} via  $1/0.021=47.62$ [$1/\mathrm{sec}$]) and $\hat{\sigma}=0.427$ [mV/$\sqrt{\mathrm{msec}}$] (it can be obtained from Table 1 in \cite{picchini2008parameters} by converting 0.0135 [V/$\sqrt{\mathrm{sec}}$] into [mV/$\sqrt{\mathrm{msec}}$]). We appreciate how close our posterior means based on 100 ISIs are to the maximum likelihood estimates using 312 ISIs.

\begin{table}[ht]
  \centering
  \small
  
  \begin{tabular}{@{}lrrrrrr@{}}
    \toprule
    Algorithm   & $\rho$ &    $N$ & CPU (m) & mESS & mESS/m  & Rel.  \\
    \midrule
    Kalman  & -  &- & 56   & 630 & 11.30 & 18.9 \\
    PMMH  & -  & 1  & 479    &287 & 0.6 & 1.0\\
    CPMMH-09  & 0.9  &  1  & 655  &  400 & 0.61 & 1.0
    \\
    CPMMH-0999  & 0.999  &  1  & 653  &  372 & 0.57 & 1.0
    \\
    \bottomrule
  \end{tabular}
  \caption{Neuronal model. Correlation $\rho$, number of particles $N$, 
    CPU time (in minutes $m$), minimum ESS (mESS), minimum ESS per minute (mESS/m), and relative minimum ESS per minute (Rel.) as compared to PMMH.  All results are based on $100$k iterations of each scheme.}\label{tab:tab_neuronal_sdemem}
\end{table}

  

\section{Discussion}
\label{sec:disc}

We have constructed an efficient and general inference methodology for the parameters of stochastic differential equation mixed-effects models (SDEMEMs). While SDEMEMs are a flexible class of models for ``population estimation'', their use has been limited by technical difficulties that make the execution of inference algorithms (both classic and Bayesian) computationally intensive. Our work proposed strategies to both (i) produce Bayesian inference for very general SDEMEMs, without the limitations of previous methods; (ii) alleviate the computational requirements induced by the generality of our methods.
The SDEMEMs we considered are general in the sense that the underlying SDEs can be nonlinear in the states and in the parameters; the random parameters can have any distribution (not restricted to the Gaussian family); the observations equation does not have to be a linear combination of the latent states. We produced a Metropolis-within-Gibbs algorithm  (hereafter Gibbs sampler, Algorithm \ref{algGibbs}) with carefully constructed blocking strategies, where the technically difficult approximation to the unavailable likelihood function is efficiently handled via correlated particle filters. The use of correlated particle filters brings in the well-known benefit of requiring fewer particles compared to the particle marginal Metropolis-Hastings (PMMH) algorithm. In our experiments, the novel blocked-Gibbs sampler embedding a correlated PMMH (CPMMH) shows that it is possible to considerably reduce the number of required particles while still obtaining a value of the effective sample size (ESS) that is comparable to using standard PMMH in the Gibbs sampler. This means that the Gibbs sampler with embedded CPMMH is computationally efficient and on two out of three examples of increasing complexity we found that our algorithm is much more efficient than a similar algorithm using the standard PMMH, sometimes even 40 times more efficient.
Some care must be taken when choosing $\rho$, which governs the level of correlation between successive likelihood estimates. Taking $\rho\approx 1$ can result in the sampler failing to adequately mix over the auxiliary variables. We found that this problem was exacerbated when using relatively few particles (such as $N=1$), but can be overcome by reducing $\rho$.
The fact that our approach is an instance of the pseudo-marginal methodology of \cite{andrieu09b} implies that we produce exact (simulation-based) Bayesian inference for the parameters of our SDEMEMs, regardless the number of particles used. 
We mostly focus on producing ``plug-and-play'' methodology (but see below for exceptions), meaning that no preliminary analytic calculations should be required to run our methods, and forward simulation from the SDEs simulator should be enough. Instead, what is necessary to set is the number of particles $N$ and, when correlated particles filters are used (CPMMH), the correlation parameter $\rho$ (however this one is easily set within the interval $[0.90,0.999]$). Finally, the usual settings for the MCMC proposal distribution should be decided (covariance matrix of the proposal function $q(\cdot)$). However, for the neuronal data example we had to employ a bridge filter, since the observational noise is very low for this case study, causing the bootstrap filter to perform poorly. The bridge filter is not plug-and-play (as discussed below), however in this paper we have decided to include a non-plug-and-play method to show how to analyze complex case studies with existing state-of-art sequential Monte Carlo filters.
When considering a plug-and-play approach, our proposed methodology relies on the use of the bootstrap particle filter, within which particles are propagated according to the SDE solution or an approximation thereof. We note that in scenarios where the observations are particularly informative (e.g. the neuronal data case study in Section \ref{sec:ou_neuronal}), it may be beneficial to propagate particles conditional on the observations, by using a carefully chosen bridge construct. We refer the reader to \cite{golightly2019} for details on the use of such constructs within a CPMMH scheme for SDEs. However, notice that in order to use the constructs in \cite{golightly2019}  the conditional distribution of observations (i.e. \eqref{eqn:obs} in our context) must be Gaussian. This is the underlying assumption that is exploited in \cite{botha2019particle} to enable the use of bridge constructs in inference for SDEMEMs. In \cite{botha2019particle} they also use methods based on correlated particle filters, in a work which has been proposed independently and concurrently to ours (July 25 2019 on arXiv). See for example their ``component-wise pseudo-marginal'' (CWPM) method, which is similar to the naive Gibbs strategy we also propose, and they found that CWPM was the best strategy among a battery of explored methods. In order to correlate the particles, \cite{botha2019particle} advocate the use of the blockwise pseudo-marginal strategy of \cite{tran2016block}: this way, at each iteration of a CPMMH algorithm they randomly pick a unit in the set $\{1,..,M\}$, and only for that unit they update the corresponding auxiliary variates, whereas for the remaining $M-1$ units they reuse the same auxiliary variates $u^i$ as employed in the last accepted likelihood approximation. This approach implies an estimated correlation between log-likelihoods of around $1-1/M$, which also implies that the correlation level is completely guided by the number of units. This means that for a small $M$ (e.g. $M=5$ or 10, implying a correlation of 0.80 and 0.90 respectively) a blockwise pseudo-marginal strategy might not be as effective as it could be. On the other hand, assuming a very efficient and scalable implementation allowing measurements from $M=10,000$ units, the blockwise pseudo-marginal approach would produce highly correlated particles, which can sometimes be detrimental by not allowing enough variety in the auxiliary variates, and ultimately producing long-term correlations in the parameter chains, as we have documented in Section \ref{sec:ou_neuronal} when using a low number of particles $N$. 
We therefore think it is advantageous to use a method that allows the statistician to decide on the amount of injected correlation: even though this means having one more parameter to set ($\rho$ in our treatment), we find this decision to be rather straightforward, as mentioned above.

We hope this work can push forward the use of SDEMEMs in applied research, as even though inference methods for SDEMEMs have been available from around 2005, the limitation of theoretical or computational possibilities have implied that only specific SDEMEMs could be efficiently handled, while other SDEMEMs needed ad-hoc solutions or computationally very intensive algorithms. We believe our work is promising as a showcase of the possibility to employ very general SDEMEMs for practical applications.   

\section*{Acknowledgments}
\noindent
SW was supported by the Swedish Research Council (Vetenskapsr{\aa}det 2013-05167). UP was supported by the Swedish Research Council (Vetenskapsr{\aa}det 2019-03924).  We thank the staff at the Center for Scientific and Technical Computing at Lund University (LUNARC) for help in setting up the computer environment used for the computations in Section  \ref{sec:ou} and \ref{sec:ou_neuronal}. We thank J. F. He for making the neuronal data available. We thank the editor and three anonymous reviewers for useful and insightful comments on this paper.

\bibliography{/bridgebib.bib}

\clearpage

\appendix

\section{Bridge particle filter} \label{sec:bridge_filter}

\subsection{Deriving the bridge filter}

This section is not strictly pertaining mixed-effects modelling, hence we disregard the subject's index. We consider the bridge particle filter proposed in \cite{golightly2011bayesian}, with the exception that there an SDE was numerically solved using the Euler-Maruyama scheme. Here we provide the bridge particle filter for the special case where the exact (Gaussian) transition density is available, as considered for case studies in Sections \ref{sec:ou} and \ref{sec:ou_neuronal}. 
Since we do not require numerical discretization, in terms of the notation established in \cite{golightly2011bayesian} we have that $m = 1$ and $j = 0$. Furthermore, we let $\Delta {\text{obs}}$ denote the step-length for the observational times grid. Thus we have that $\Delta t = \Delta {\text{obs}}$ and $\Delta {j=0} = \Delta {\text{obs}}$. 

Here the bridge filter is derived for the example in section \ref{sec:ou_neuronal}.
The analytical transition density for the $X_t$ process in \eqref{sec:ou_neuronal} is 

\begin{align*}
    X_{t+\Delta t} | X_t = x_ t \sim \textrm{N}\biggl(x_t e^{-\lambda\Delta t} + \frac{\nu}{\lambda}(1-e^{-\lambda\Delta t}), \frac{\sigma^2}{2\lambda}(1-e^{-2\lambda\Delta t})\biggr).
\end{align*}
The joint density for $X_{t+\Delta t}$ and $Y_{t+\Delta t}$, conditional on $X_t$, is 

\begin{align*}
    \begin{pmatrix}
    X_{t+\Delta t} \\ Y_{t+\Delta t}
    \end{pmatrix} | X_t = x_ t \sim \textrm{N}\Big \{   \begin{pmatrix}
    \alpha_0  \\ \alpha_0 
    \end{pmatrix} ,  \begin{pmatrix}
    \beta_0 & \beta_0 \\ \beta_0 & \beta_0 + \sigma^2_{\epsilon}
    \end{pmatrix} \Big \}
\end{align*}
where $\alpha_0 = x_t e^{-\lambda\Delta t} + \frac{\nu}{\lambda}(1-e^{-\lambda\Delta t})$, and $\beta_0 = \frac{\sigma^2}{2\lambda}(1-e^{-2\lambda\Delta t})$. The conditional distribution used as proposal distribution in the bridge filter is 

\begin{align}\label{eq:cond_trans_dist}
    \hat{\pi}(x_{t + \Delta t} | x_{t}, y_{t + \Delta t}) = \textrm{N}(x_{t + \Delta t}; \mu, \Sigma),
\end{align}
where $\mu = \alpha_0 + \beta_0(\beta_0  + \sigma^2_{\epsilon})^{-1}(y_{t + \Delta t}- \alpha_0)$, $\Sigma = \beta_0 (1 - [ \beta_0 + \sigma^2_{\epsilon}  ]^{-1}\beta_0)$. 

Equation \eqref{eq:cond_trans_dist} can be used to propagate particles forward, which is a much more efficient approach than in the bootstrap filter case, where the sampler is miopic to the next observation, while \eqref{eq:cond_trans_dist} is able to look-ahead towards the next observation $y_{t+\Delta t}$. Thus, the bridge filter is similar in structure to Algorithm \ref{BPF} with the difference that here the particles propagation step consists in sampling from \eqref{eq:cond_trans_dist}, and the weights are given by

\begin{align*}
    \tilde{w}_{t + \Delta t,k}= \frac{\pi(y_{t+\Delta t}|x_{t+\Delta t,k},\sigma_{\epsilon}^2) \pi(x_{t+\Delta t,k} |x_{t,k} ) }{\hat{\pi}(x_{t+\Delta t,k} |x_{t,k},y_{t+\Delta t})}, \qquad  w_{t + \Delta t,k}=\frac{ \tilde{w}_{t + \Delta t,k}}{\sum_{j=1}^{N} \tilde{w}_{t + \Delta t,j}}, \qquad k=1,...,N.
\end{align*}

\subsection{Comparing the bootstrap filter and the bridge particle filter}  \label{appendix:comp_bootstrap}

\begin{figure}[ht]
\centering
\includegraphics[scale = 0.6]{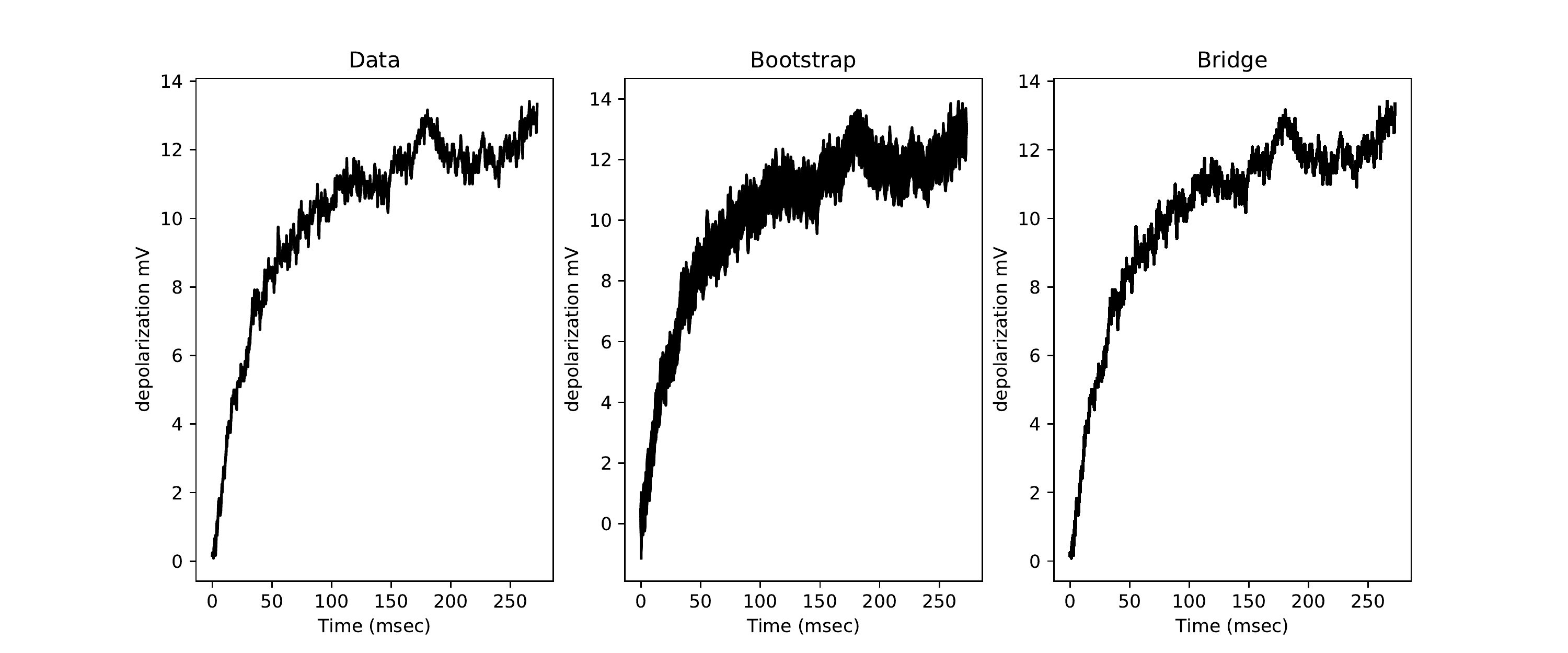}
\caption{Neuronal model: forward propagation of the particles for bootstrap and bridge filter for one ISI (chosen at random; this ISI contained 1817 data points). Leftmost panel: observed data for that ISI. Central panel: forward propagation of the particles from the bootstrap filter. Rightmost panel: forward propagation of the particles from the bridge filter.}
\label{fig:bootstrap_bridge_tracking}
\end{figure}

To compare the performance of the bootstrap and the bridge filter, we run both filters with the same number of particles (500 particles for each subject) using the 100 ISIs neuronal data from Section \ref{sec:ou_neuronal}. Parameters are set at the posterior means obtained from the Kalman algorithm. The comparison is interesting since it illustrates the well known issue of running particle filters when the observational error is small (here we have that $\sigma_{\epsilon} \approx 0.001$), and hence it is expected that the bootstrap filter will produce sub-optimal results. This is due to its inability to ``target'' the next observation, thus producing very small weights due to the small $\sigma_\epsilon$.
In Figure \ref{fig:bootstrap_bridge_tracking}, we compare the forward propagation of the particles for one ISI chosen at random. It is evident that the bridge filter follows the data more closely. Furthermore, we run each filter independently for 100 times and compare the averages of the log-likelihood values, the standard deviation of the 100 log-likelihood estimations, and the runtimes, see Table \ref{tab:bootstrap_bridge_tracking}. We can easily notice the superiority of the bridge filter returning an averaged log-likelihood value very close to the one provided by the Kalman filter. In particular, notice how the log-likelihood estimation is very unreliable (due to the small observation error). 

\begin{table}[ht] 
\centering
\caption{Comparing 100 log-likelihood estimations for the bootstrap and bridge filter. }
\label{tab:bootstrap_bridge_tracking}
\begin{tabular}{lrrr}
\toprule
          & Log-likelihood  & Std. Dev.   & Runtime (sec) \\ 
\midrule
Kalman    & 62091   & -    & 0.012  \\ 
Bootstrap & -2594152 & 119905 &  21.51  \\ 
Bridge    & 62291   & 0.34  & 27.50  \\ 
\bottomrule
\end{tabular}
\end{table}

\begin{figure}[h]
\centering
\includegraphics[scale=0.6]{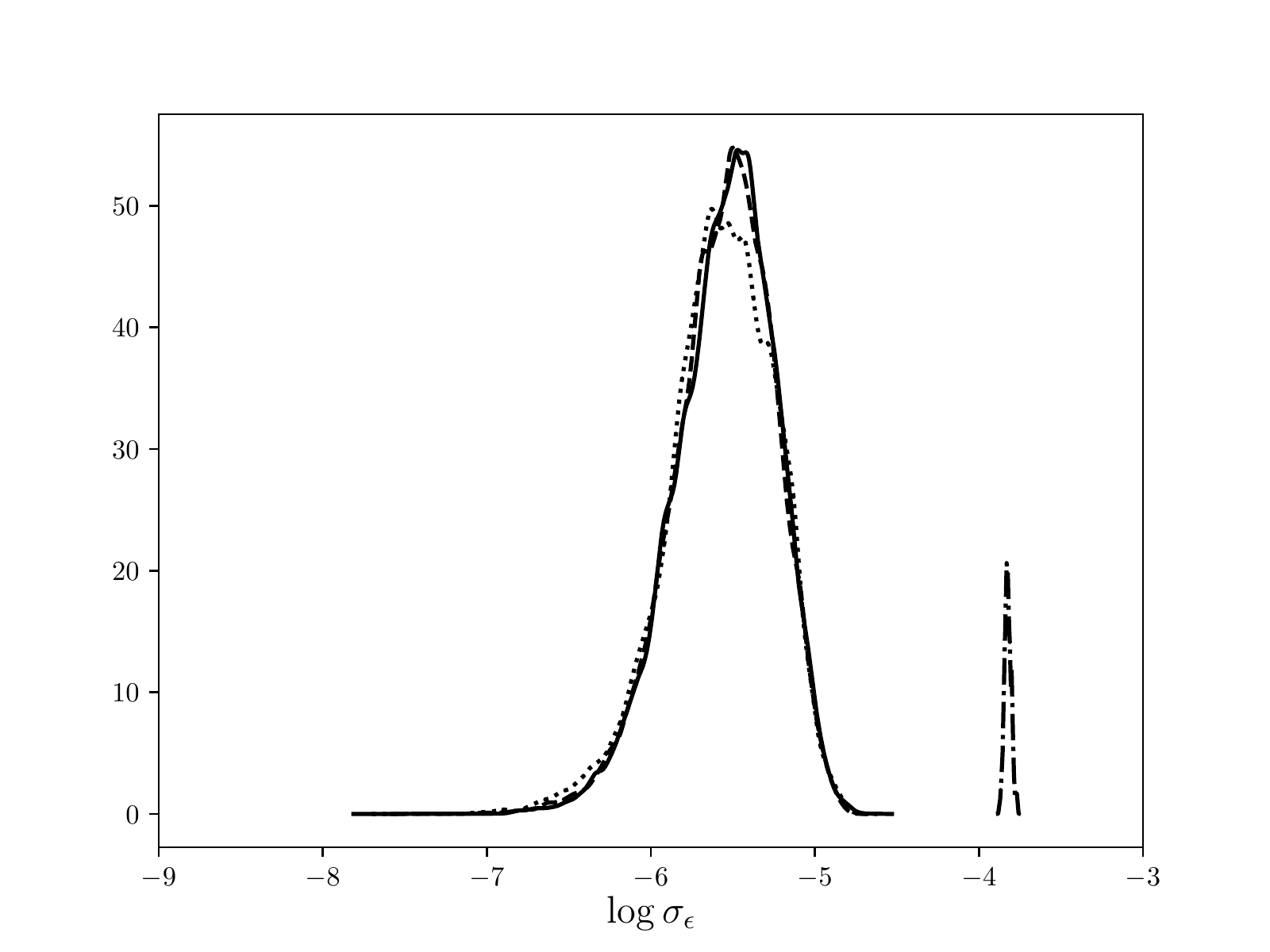}
\caption{Neuronal model: marginal posterior distributions for $\log\sigma_{\epsilon}$. Solid line is Kalman, dashed line is PMMH using the bridge filter, dotted line is CPMMH-09 using the bridge filter, dash-dotted line is CPMMH-099 using the bootstrap filter. The marginal posteriors for Kalman, PMMH, and CPMMH-09 have been multiplied by a factor 40 for pictorial reasons.}
\label{fig:neuronal_mp_sigma_epsilon_appendix}
\end{figure}

We now compare the inference results for CPMMH when using the bridge filter and the bootstrap filter. We ran four algorithms: Kalman, PMMH with N = 1 particles using the bridge filter, CPMMH-09 with N = 1 particles using the bridge filter, CPMMH-099 with N = 100 particles using the bootstrap filter. We ran, Kalman, PMMH, and CPMMH-09  for 100k iterations, and ran CPMMH-099 for only 35k iterations, as this case is computationally more intensive. 
In Figure \ref{fig:neuronal_mp_sigma_epsilon_appendix} we see that when using the bootstrap filter driven inference scheme, the $\sigma_{\epsilon}$ chain fails to adequately explore regions of high posterior density. We emphasise that this is due to using too few particles ($N=100$). It is clear from Table~\ref{tab:bootstrap_bridge_tracking} that the number of particles required to match the efficiency of the bridge filter is computationally infeasible.
Marginal posteriors for the remaining parameters (not shown) are however similar for all algorithms.  The reason why the population parameters $\eta$ appear to be unaffected by these issues, unlike $\sigma_\epsilon$, is that step 4 of the Gibbs algorithms in section \ref{sec:gibbs} (both versions, naive and blocked one) does not depend on the approximated likelihood, whereas step 2 (which samples $\sigma_\epsilon$) does depend on it.

\section{Tumor growth -- Linear noise approximation}\label{sec:lna}
The linear noise approximation (LNA) can be derived in a number of more or less formal ways. We present a brief informal derivation here and refer the reader to \cite{Fearnhead2014} and the references therein for further details. We remark that the LNA is not a necessary feature of our general plug-and-play methodology outlined in Section \ref{sec:pmmh} and Algorithm \ref{algGibbs}. 

\subsection{Setup}
Consider the tumor growth model in \eqref{model}, \eqref{obsmodel} and \eqref{randmodel} and a single experimental unit so that the superscript $i$ can be dropped from the notation. To obtain a tractable observed data likelihood, we construct the linear noise approximation of $\log V_t = \log(X_{1,t}+X_{2,t})$.

Let $Z_t=(Z_{1,t},Z_{2,t},Z_{3,t})^T=(\log V_t,\log X_{1,t},\log X_{2,t})^T$. The SDE satisfied by $Z_t$ can be found using the It\^o formula, for which we obtain
\[
dZ_t = \alpha(Z_t,\phi)dt+\sqrt{\beta(Z_t,\phi)}dW_t
\] 
where
\[
\alpha(Z_t,\phi)=\begin{pmatrix}
\left\{\beta+0.5\gamma^2\right\}e^{Z_{2,t}-Z_{1,t}}+\left\{-\delta+0.5\tau^2\right\}e^{Z_{3,t}-Z_{1,t}}-0.5\left\{\gamma^2e^{2(Z_{2,t}-Z_{1,t})}+\psi^2e^{2(Z_{3,t}-Z_{1,t})}\right\} \\
\beta\\
-\delta
\end{pmatrix}
\]
\[
\beta(Z_t,\phi)=\begin{pmatrix}
\gamma^2e^{2(Z_{2,t}-Z_{1,t})}+\tau^2e^{2(Z_{3,t}-X_{1,t})} & \gamma^2e^{2(Z_{2,t}-Z_{1,t})} & \psi^2e^{2(Z_{3,t}-Z_{1,t})}\\
\gamma^2e^{2(Z_{2,t}-Z_{1,t})} & \gamma^2 & 0 \\
\psi^2e^{2(Z_{3,t}-Z_{1,t})} & 0 & \psi^2
\end{pmatrix}.
\]
We apply the linear noise approximation (LNA) by partitioning $Z_t$ as $Z_t=m_t+R_t$ where 
$m_t$ is a deterministic process satisfying
\begin{equation}\label{ode1}
\frac{dm_t}{dt}=\alpha(m_t,\phi)
\end{equation}
and $\{R_t,t\geq 0\}$ is a residual stochastic process satisfying
\[
dR_t=\left\{\alpha(Z_t,\phi)-\alpha(m_t,\phi)\right\}dt+\sqrt{\beta(Z_t,\phi)}dW_t.
\]
By Taylor expanding $\alpha$ and $\beta$ about the deterministic process $m_t$ and retaining the first two terms in the expansion of $\alpha$, and the first term in the expansion of $\beta$, we obtain an approximate residual stochastic 
process $\{\tilde{R}_t,t\geq 0\}$ satisfying
\[
d\tilde{R}_t=J_t\tilde{R}_t dt+\sqrt{\beta(m_t,\phi)}dW_t
\]
where $J_t$ is the Jacobian matrix with $(i,j)$th element 
$(J_t)_{i,j}=\partial\alpha_i(m_t,\phi)/\partial m_{j,t}$. Assuming initial values $m_0=z_0$ and $\tilde{R}_0=0$, the approximating distribution of $Z_t$ is given by  
\begin{equation}\label{LNA}
Z_t|Z_0=z_0 \approx \textrm{N}(m_t,H_t)
\end{equation}
where $m_t$ satisfies (\ref{ode1}) and, after several calculations which we omit for brevity, $H_t$ is the solution to 
\begin{equation}\label{ode2}
\frac{dH_{t}}{dt}=H_t J_t^T+\beta(m_t,\phi)+J_t H_t.
\end{equation}

\subsection{Inference}
Note that the observation model in (\ref{obsmodel}) can be written as
\begin{equation}\label{obsmodel2}
Y_t=P^T Z_t + \epsilon_t,\qquad \epsilon_t\indep \textrm{N}(0,\sigma^2_e).
\end{equation}
where $P$ is a $3\times 1$ `observation vector' with first entry 1 and zeroes elsewhere. The linearity of (\ref{LNA}) and (\ref{obsmodel2}) yields 
a tractable approximation to the marginal likelihood $\pi(y|\phi,\sigma_e)$, which we denote by $\pi_{\textrm{LNA}}(y|\phi,\sigma_e)$. The approximate marginal likelihood $\pi_{\textrm{LNA}}(y|\phi,\sigma_e)$ can be factorised as 
\begin{equation}
\pi_{\textrm{LNA}}(y|\phi,\sigma_e)=\pi_{\textrm{LNA}}(y_{1}|\phi,\sigma_e)\prod_{i=2}^{n}\pi_{\textrm{LNA}}(y_{i}|y_{1:i-1},\phi,\sigma_e)
\end{equation}
where $y_{1:i-1}=(y_{1},\ldots,y_{i-1})^T$. Suppose that 
$Z_{1}\sim \textrm{N}(a,C)$ \emph{a priori}, for some constants $a$ and $C$. The marginal likelihood 
under the LNA, $\pi_{\textrm{LNA}}(y_{1:n}|\phi,\sigma_e):=\pi_{\textrm{LNA}}(y|\phi,\sigma_e)$ can be obtained via a forward filter, which is given in Algorithm~\ref{algFF}.
\begin{algorithm}[ht]
\footnotesize
\caption{Forward filter}\label{algFF}
\textbf{Input:} Data $y$, parameter values $\phi$ and $\sigma_e$.\\
\textbf{Output:} Observed data likelihood $\pi_{\textrm{LNA}}(y|\phi,\sigma_e)$.
\begin{enumerate}
\item Initialisation. Compute 
\[
\pi_{\textrm{LNA}}(y_{1}|\phi,\sigma_e)=N\left(y_{1}\,;\, P^T a\,,\,P^T C P+\sigma^2_e\right)
\]
where $\textrm{N}(\cdot\,;\,a\,,\,C)$ denotes the Gaussian density with 
mean vector $a$ and variance matrix $C$. The posterior at time $t=1$ is therefore 
$Z_{1}|y_{1}\sim \textrm{N}(a_{1},C_{1})$ where
\begin{align*}
a_{1} &= a+C P\left(P^T C P+\sigma^2_e\right)^{-1}\left(y_{1}-P^T a\right) \\
C_{1} &= C-C P\left(P^T C P+\sigma^2_e\right)^{-1}P^T C\,.
\end{align*}
 
\item For $i=1,2,\ldots ,n-1$,
\begin{itemize}
\item[(a)] Prior at $i+1$. Initialise the LNA with $m_{i}=a_{i}$ and $H_{i}=C_{i}$. 
Integrate the ODEs (\ref{ode1}) and (\ref{ode2}) 
forward to $i+1$ to obtain $m_{i+1}$ and $H_{i+1}$. Hence
\[
Z_{i+1}|y_{1:i}\sim \textrm{N}(m_{i+1},H_{i+1})\,.
\]
\item[(b)] One step forecast. Using the observation equation, we have that 
\[
Y_{i+1}|y_{1:i}\sim N\left(P^T m_{i+1},P^T H_{i+1}P+\sigma^2_e\right)\,.
\]
Compute
\begin{align*}
\pi_{\textrm{LNA}}(y_{1:i+1}|\phi,\sigma_e)&=\pi_{\textrm{LNA}}(y_{1:i}|\phi,\sigma_e)\pi_{\textrm{LNA}}(y_{i+1}|y_{1:i},\phi,\sigma_e)\\
&=\pi_{\textrm{LNA}}(y_{1:i}|\phi,\sigma_e)\,N\left(y_{i+1}\,;\, P^T m_{i+1}\,,\,P^T H_{i+1}P+\sigma^2_e\right)\,.
\end{align*}
\item[(c)] Posterior at $i+1$. Combining the distributions in (a) and (b) gives the joint 
distribution of $Z_{i+1}$ and $Y_{i+1}$ (conditional on $y_{1:i}$ and $\phi$) as
\[
\left(\begin{array}{c}
	Z_{i+1} \\	
	Y_{i+1}
	\end{array}\right)\sim N\left\{\left(\begin{array}{c}
	m_{i+1} \\
	P^T m_{i+1} 	
	\end{array} \right )\,,\, \left(\begin{array}{cc}
	H_{i+1} & H_{i+1}P  \\
	P^T H_{i+1} & P^T H_{i+1}P+\sigma^2_e  	 
	\end{array}   \right ) \right \} 
\]
and therefore $Z_{i+1}|y_{1:i+1}\sim \textrm{N}(a_{i+1},C_{i+1})$ where
\begin{align*}
a_{i+1} &= m_{i+1}+H_{i+1}P\left(P^T H_{i+1}P+\sigma^2_e\right)^{-1}\left(y_{i+1}-P^T m_{i+1}\right) \\
C_{i+1} &= H_{i+1}-H_{i+1}P\left(P^T H_{i+1}P+\sigma^2_e\right)^{-1}P^T H_{i+1}\,.
\end{align*}

\end{itemize}
\end{enumerate}
\end{algorithm}   

Inference for the SDEMEM defined by (\ref{model}), (\ref{obsmodel}) and (\ref{randmodel}) may be performed via a Gibbs sampler that draws from the following full conditionals
\begin{enumerate}
\item $\pi_{\mathrm{LNA}}(\phi|\eta,\sigma_e,y)\propto \prod_{i=1}^M \pi(\phi^i|\eta)\pi_{\textrm{LNA}}(y^i|\sigma_e,\phi^i)$,
\item $\pi_{\mathrm{LNA}}(\sigma_e|\eta,\phi,y)\propto \pi(\sigma_e)\prod_{i=1}^M \pi_{\textrm{LNA}}(y^i|\sigma_e,\phi^i) $,
\item $\pi(\eta|\sigma_e,\phi,y)\propto \pi(\eta)\prod_{i=1}^M \pi(\phi^i|\eta)$. 
\end{enumerate}

\end{document}